\documentclass[twocolumn,caps,showpacs,superscriptaddress]{revtex4-1}

\usepackage{graphicx,color,soul}
\usepackage{latexsym}
\usepackage{amsmath,amssymb}        
\usepackage[draft=false]{hyperref}

\begin{document}


\title{Spherical non-linear absorption of cosmological scalar fields onto a black hole}


\author{F. S. Guzm\'an}
\affiliation{Instituto de F\'{\i}sica y Matem\'{a}ticas, Universidad
              Michoacana de San Nicol\'as de Hidalgo. Edificio C-3, Cd.
              Universitaria, 58040 Morelia, Michoac\'{a}n,
              M\'{e}xico.}

\author{F. D. Lora-Clavijo}
\affiliation{Instituto de F\'{\i}sica y Matem\'{a}ticas, Universidad
              Michoacana de San Nicol\'as de Hidalgo. Edificio C-3, Cd.
              Universitaria, 58040 Morelia, Michoac\'{a}n,
              M\'{e}xico.}


\date{\today}


\begin{abstract}
In this paper we track the non-linear spherical evolution of a massless scalar field onto a Schwarzschild black hole space-time as a first approximation to the accretion of cosmologically motivated classical scalar fields. We perform an analysis related to wave packets described by wave number and width. We study various values of the wave number and found that for $k=0$ and width packets bigger than the Schwarzschild radius, the absorption is not total. In the cases we studied for $k>0$, the black hole absorbs the total amount of energy density of the scalar field moving toward the horizon. Our results indicate that assuming spherical symmetry, in the non-linear regime, there are cases for which scalar fields are allowed to survive outside black holes and may eventually have life-times consistent with cosmological time scales.
\end{abstract}




\maketitle


\section{Introduction}

Cosmological scalar fields have been an important ingredient of cosmological models since the model of inflation proposed a mechanism providing exponential cosmological growth \cite{guth}. Considering the supernovae redshift observations a smooth classical scalar field was proposed to play the role of dark energy \cite{quintessence}.  And scalar fields have also been proposed to play the role of dark matter, first at galactic scale \cite{sfdm-gal} and then at cosmic scale \cite{sfdm-cosmo}. On the other hand, current observations indicate that there are supermassive black holes in the centers of galaxies  with masses of the order of $10^6 - 10^9$ solar masses \cite{smbh}. One of the quests consists in determining the components feeding the black hole, mainly related to the accretion of either  baryons or dark matter. For instance it can be considered that supermassive black holes grow through the accretion of baryonic and dark matter onto seed black holes of intermediate mass of about $10^3$ to $10^4 M_{\odot}$  \cite{seedBHs}. Also some results indicate that only about 10 \% of their mass is due to the accretion of dark matter (e.g. \cite{freitas}), and more mature models involving the study of the Newtonian phase space \cite{losecone} indicate that the time-scale  for the accretion of collisionless dark matter to contribute significantly to a supermassive black hole mass is too long (see e.g. \cite{Gilmore}). Currently the study of ideal gas dark matter accretion onto supermassive black holes is under research and important bounds on the equation of state of dark matter may arise \cite{GuzmanLora2011, GuzmanLora2011b}.

Thus studying the accretion of cosmologically motivated scalar fields is expected to have an astrophysical impact within the subject of the viability of dark matter candidates. In fact the accretion of scalar field dark matter and dark energy onto supermassive black holes has been already explored under certain assumptions. In \cite{LiddleUrena} the accretion rate of scalar fields is calculated in order to estimate whether or not the accretion rate is consistent with the mass and life time scale of supermassive black holes. Later on, in \cite{CruzGuzmanLora} the same problem was treated considering that the scalar field has two ends, the black hole's horizon and future null infinity, which was achieved by using hyperboloidal slices of space-time and compact radial coordinate \cite{Anil} which allowed the scalar field not only to be accreted by the black hole, but also to leak through scri+, and it was found an extremely high dilution rate of the scalar field density that may rule our scalar fields as dark matter and dark energy. On the other hand, more recently in \cite{UrenaFernandez} it was studied the accretion rate in terms of the wave number of a wave packet of scalar field, and found interesting results related to dependence of accretion rates on the width and wave number of wave packets, that is, for wave packets with small wave number and very spread the absorption of the scalar field is lower than for thin packets with big wave numbers; this problem was also studied under a different context related to the absorption of light in \cite{mendozza}. In \cite{unam} the same problem was treated and it was found that there are particular initial scalar field profiles that allow long-lived scalar field densities around a black hole. All these results have been obtained considering spherical symmetry and that the scalar field is a test field, that is, the geometry of the space-time remains fixed. 

In this paper explore a step forward and include the evolution of the space-time geometry. That is, we  study a spherically symmetric scalar field evolving onto an existing black hole space-time, considering the evolution of the geometry too, so that we can measure the black hole's horizon growth and determine more precise accretion rates and absorption ratios. We focus on the study of the absorption of the scalar field in terms of the parameters of wave number and width of the initial wave packet as described in \cite{UrenaFernandez}. 

In order to clearly set the astrophysical scenario we deal with, we consider important to stress that the black hole is already formed initially, and that it is asymptotically flat. These conditions have important implications: 
(i) we assume the cosmological effects on the system are neglected, and the evolution of the system is ruled by the gravity due to the black hole and the scalar field only; that is, we assume our system occurs much after the early universe stage, unlike previous studies related to primordial black holes growth through the accretion of a scalar field \cite{Harada};
(ii) in most of cosmological models involving scalar fields, a first approximation of homogeneity and isotropy is assumed \cite{quintessence,sfdm-cosmo}, however at local scales -for instance close to the event horizon of a black hole- we consider such conditions do not hold.
(iii) sometimes properties of pressure an density, and thus an equation of state, are attached to cosmological scalar fields \cite{quintessence,Harada}, however in general at local scales scalar fields develop anisotropic pressures and time and space dependent relations between density and pressure (see \cite{oscillatons} for scalar field self-gravitating configurations and \cite{phantom} for the accretion of a phantom scalar field onto a black hole) that we consider irrelevant in our analysis.

This paper is organized as follows. In Sec. \ref{sec:equations} we describe the system of equations ruling the evolution of the scalar field and the geometry, in Sec. \ref{sec:NM} we describe the numerical methods used to solve the equations. In sec \ref{sec:results} we describe our results and in Sec. \ref{sec:conclusions} we discuss our results and draw some conclusions.

\section{Evolution equations} 
\label{sec:equations}

\subsection{Evolution of the geometry}

In order to solve numerically the Einstein Field equations $G_{\mu \nu}=8\pi T_{\mu\nu}$, where $G_{\mu\nu}$ is the Einstein tensor and $T_{\mu \nu}$ is the energy momentum tensor, we use the 3+1 decomposition space-time and adopt the ADM system of evolution equations. We start up with the metric 

\begin{equation}
ds^2 = -(\alpha^2 - \beta_i \beta^i)dt^2 + 2\beta_i dx^i dt + \gamma_{ij}dx^i dx^j, \label{eq:lineelement}
\end{equation}

\noindent where $\alpha$ is the lapse function, $\beta^i$ are the components of the shift vector, $\gamma_{ij}$ are the components of the 3-metric of the hyper-surfaces that foliate the space-time and  $x^\mu=(t,x^i)$ are the coordinates of the space-time. All our calculations assume geometric units $G=c=1$.

According to the ADM formulation of general relativity, Einstein's equations split into evolution equations for the 3-metric $\gamma_{ij}$ and the extrinsic curvature $K_{ij}$ of the hypersurfaces 

\begin{eqnarray}
\partial_t \gamma_{ij} &=& -2\alpha K_{ij} + \nabla_i \beta_j + \nabla_j 	\beta_i, \label{eq:gammaevolve}\\
\partial_t K_{ij} = &-& \nabla_i \nabla_j \alpha + \alpha \left (  R_{ij} + K K_{ij} - 2K_{il}K^l_j  \right ) \nonumber\\
&+& 4\pi\alpha \left [ (S - \rho_{ADM})\gamma_{ij}  - 2S_{ij} \right] \nonumber\\
&+& \beta^l\nabla_l K_{ij} + K_{il}\nabla_j \beta^l + K_{jl}\nabla_i \beta^l, \label{eq:Kevolve}
\end{eqnarray}

\noindent where $\nabla_i$ denotes the covariant derivative with respect to the 3-metric, $R_{ij}$ is the Ricci tensor related to the spacelike hypersurfaces and $K=\gamma^{ij}K_{ij}$ is the trace of the extrinsic curvature. In addition to the evolution equations, there are the Hamiltonian and Momentum constraints

\begin{eqnarray}
^{(3)}R + K^2 - K_{ij}K^{ij} - 16\pi\rho_{ADM} = 0, \nonumber\\
\nabla_j K^{ij} - \gamma^{ij} \nabla_j K - 8\pi j^i = 0, \label{eq:Mom} 
\end{eqnarray}

\noindent where $^{(3)}R$ is the scalar of curvature associated to $\gamma_{ij}$. In the equations (\ref{eq:gammaevolve} - \ref{eq:Mom}), the quantities $\rho_{ADM}$, $j^i$, $S_{ij}$ and $S=\gamma^{ij}S_{ij}$ correspond to the local energy density, the momentum density, the spatial stress tensor and its trace respectively, measured by an Eulerian observer. These variables are obtained from the projection of the energy momentum tensor  $T_{\mu \nu}$ along the spacelike hypersurfaces and along the normal direction to such hypersurfaces.
 
We will now restrict to spherically symmetric black holes. In such case we consider the following ansatz  for the 3-metric  $\gamma_{ij}$, the extrinsic curvature $K_{ij}$ and the shift vector $\beta^i$:

\begin{subequations}
  \begin{equation}
     \gamma_{ij} = \left( 
         \begin{array}{ccc}
           \gamma_{rr} & 0 & 0 \\
           0 & \gamma_{\theta \theta} & 0 \\
           0 & 0 & \gamma_{\theta \theta}  \sin^2 \theta 
         \end{array}
     \right),
  \end{equation}
   \begin{equation}
      K_{ij} = \left( 
         \begin{array}{ccc}
           K_{rr} & 0 & 0 \\
           0 & K_{\theta \theta} & 0 \\
           0 & 0 & K_{\theta \theta}  \sin^2 \theta 
         \end{array}
     \right),
  \end{equation}
   \begin{equation}
     \beta^{i} = \left ( 
         \beta^r, 0, 0
     \right ),
  \end{equation}
\end{subequations}

\noindent where the usual polar spherical topology with spatial coordinates $x^i=(r,\theta,\varphi)$ is used. The evolution of the space-time geometry described in general by (\ref{eq:gammaevolve}) and (\ref{eq:Kevolve}) reduces in the present case to the following set of equations

\begin{eqnarray}
\partial_t \gamma_{rr} &=& -2\alpha K_{rr} + \beta^r \partial_r \gamma_{rr} + 2\gamma_{rr} \partial_r 		\beta^r, \nonumber \\ 
\partial_t \gamma_{\theta \theta} &=& -2\alpha K_{\theta \theta} + \beta^r \partial_r \gamma_{\theta \theta}, 	\nonumber \\
\partial_t K_{rr} &=& -\partial_{rr}\alpha + \frac{(\partial_r \gamma_{rr})( \partial_r \alpha)}{2\gamma_{rr}} 		+ \frac{\alpha}{2}\left (\frac{\partial_r \gamma_{\theta \theta}}{\gamma_{\theta \theta}}\right )^2 
          \nonumber \\
&-& \alpha \frac{\partial_{rr} \gamma_{\theta \theta}}{\gamma_{\theta \theta}} + \alpha \frac{(\partial_r 		\gamma_{rr})(\partial_r \gamma_{\theta \theta})}{2\gamma_{rr} \gamma_{\theta \theta}} + 2\alpha		\frac{K_{rr} K_{\theta \theta}}{\gamma_{\theta \theta}} \nonumber\\ 
&-& \alpha \frac{K_{rr}^2 }{\gamma_{rr}} + \beta^r\partial_r K_{rr} + 2K_{rr}\partial_r \beta^r \nonumber\\ 
&+& 4\pi \alpha [(S-\rho_{ADM})\gamma_{rr} - 2S_{rr}], \nonumber \\  
\partial_t K_{\theta \theta} &=& -\frac{(\partial_r \gamma_{\theta \theta})(\partial_r \alpha)}{2\gamma_{rr}} - 	\alpha \frac{\partial_{rr} \gamma_{\theta \theta}}{2\gamma_{rr}} \nonumber \\ 
&+& \alpha \frac{(\partial_r \gamma_{rr})(\partial_r \gamma_{\theta \theta})}{4\gamma_{rr}^2} + \alpha 		\left [ 1 + \frac{K_{rr} k_{\theta \theta}}{\gamma_{rr}} \right ]  + \beta^r \partial_r K_{\theta \theta} 			\nonumber\\
 &+& 4\pi \alpha [(S-\rho_{ADM})\gamma_{\theta \theta} - 2S_{\theta \theta}]\label{eq:admspherical}.
\end{eqnarray}  

These are the equations of the evolution of geometry that have to be evolved together with the matter field equations.

\subsection{Evolution of the scalar field} 

The matter field in the equations above corresponds to a scalar field described by the stress-energy tensor

\begin{equation}
T_{\mu\nu} = \Phi_{,\mu}\Phi_{,\nu} - \frac{1}{2}g_{\mu\nu}[\Phi^{,\delta}\Phi_{,\delta} + m^2 \Phi^2],
\end{equation}

\noindent whose evolution equation is given by the Bianchi identities $T^{\mu\nu}{}_{;\nu}=0$, which reduces to the Klein-Gordon equation 

\begin{equation}
\Box \Phi - m^2 \Phi = 0, \label{eq:kg}
\end{equation}

\noindent where the D'Alambertian operator for a general space-time is $\Box \Phi = \frac{1}{\sqrt{-g}}\partial_{\mu}[\sqrt{-g}g^{\mu\nu}\partial_{\nu}\Phi]$. Like in \cite{UrenaFernandez}, we restrict to the mass-less case as a first approximation of the full study of the parameter space of the scalar field properties. In order to couple the evolution of the geometry and matter with the same evolution algorithm we write equation (\ref{eq:kg}) for $m=0$ as a first order system of equations as suggested in \cite{Jonathan}:

\begin{eqnarray}
\partial_t \Pi &=& \partial_r (\alpha \gamma^{rr} \Psi) + \alpha K \Pi + \beta^r \partial_r \Pi,\nonumber\\
\partial_t \Psi &=& \partial_r (\alpha \Pi + \beta^r \Psi), \nonumber\\
\partial_t \Phi &=& \alpha \Pi + \beta^r \Psi,\label{eq:ke1st}
\end{eqnarray}

\noindent where $\Pi = (\partial_t \Phi - \beta^r \partial_r \Phi )/\alpha$ and $\Psi = \partial_r \Phi$ are new first order variables. In terms of these new variables, the source terms in the ADM equations read:

\begin{eqnarray}
4\pi \rho_{ADM} &=& \frac{1}{2}\left( \gamma^{rr} \Psi^2 + \Pi^2 \right),\nonumber\\
4 \pi j_r &=& - \Psi \Pi, \nonumber\\
4\pi S_{rr} &=& \Psi^2 + \frac{1}{2}\gamma_{rr} (-\gamma^{rr}\Psi^2 + \Pi^2),\nonumber\\
4\pi S &=& -\frac{1}{2}\gamma^{rr}\Psi^2 + \frac{3}{2}\Pi^2.\label{eq:physicalquantities}
\end{eqnarray}

\noindent In summary, the evolution equations to be solved are (\ref{eq:admspherical}) and (\ref{eq:ke1st}) subject to the constraints (\ref{eq:Mom}).

\section{Numerical Methods}
\label{sec:NM}

We solve the system of equations as an initial value problem using a finite differences approximation on a single resolution uniform grid. A description of each stage of the solution is as follows.

\subsection{Initial Data}

In order to start up the evolution it is necessary to construct initial data consistent with Einstein's equations, unlike the cases where the scalar filed is a test field \cite{CruzGuzmanLora,UrenaFernandez,unam,mendozza}, where arbitrary initial scalar field profiles are used on top of a given background. 
We simplify the constraints (\ref{eq:Mom}) by assuming the ansatz $\gamma_{\theta\theta} = r^2$ and $K_{rr}=-\frac{2M}{r^2}\frac{1+M/r}{\sqrt{1+2M/r}}$, which reduce the constraints (\ref{eq:Mom}) to the following set of ordinary equations

\begin{eqnarray}
\partial_r \gamma_{rr} &=& \frac{\gamma_{rr}}{r}\left(1 - \gamma_{rr} - \frac{\gamma_{rr} K_{\theta\theta}^{2}}{r^2} \right) - 2\frac{\gamma_{rr}}{r}K_{rr}K_{\theta\theta} \nonumber\\
&+& 8\pi r \gamma_{rr}^{2} \rho_{ADM}, \nonumber\\
\partial_r K_{\theta\theta} &=& r\frac{K_{rr}}{K_{\theta\theta}} + \frac{K_{\theta\theta}}{r} - 4\pi r^2 \gamma_{rr}j^r.\label{eq:specialconstraints}
\end{eqnarray}

\noindent We solve these equations once we have the explicit matter fields, for which we need to provide a scalar field profile similar to that described in \cite{UrenaFernandez} which corresponds to a  spherical wave modulated by a Gaussian profile:

\begin{eqnarray}
\Phi(t=0,r) &=& A \frac{\cos(kr)}{r} e^{-(r-r_0)^2/\sigma^2}, \nonumber\\
\Psi(t=0,r) &=& \frac{\partial \Phi(t=0,r)}{\partial r},\nonumber\\
\Pi(t=0,r) &=& -\frac{\beta^r}{\alpha}\Psi(t=0,r).\label{eq:idSF}
\end{eqnarray}

\noindent With this information we calculate $\rho_{ADM}$ and $j^r $ at initial time using (\ref{eq:physicalquantities}), then substitute into equations (\ref{eq:specialconstraints}) and solve using a fourth order Runge-Kutta integrator.

Unlike the problems where the space-time is assumed to remain fixed (e. g. \cite{CruzGuzmanLora,UrenaFernandez,unam}) the space-time is initially distorted with respect to an exact black hole geometry. This fact makes difficult to filter ingoing and outgoing pure modes of the scalar field initially as described in \cite{UrenaFernandez}. Therefore we do not attempt to have only ingoing or outgoing pure modes and will keep the presence of both. Since we want a mass of reference to compare with the growth of the black hole's horizon, we calculate the Misner-Sharp mass (see below) which contains the information -in terms of mass- of the ingoing and outgoing pulses. We will thus focus on the Misner-Sharp mass related to the ingoing pulse and define the ADM mass as the Misner-Sharp mass of the space-time minus the mass carried by the outgoing pulse.

\subsection{Evolution}

In order to solve the evolution equations (\ref{eq:admspherical}) and (\ref{eq:ke1st}) we use the method of lines with a third order Runge-Kutta integrator and second order accurate spatial stencils. On the other hand, Bianchi identities guarantee that if the Hamiltonian and momentum constraints are satisfied at initial time, they would be satisfied further. We monitor that the constraints are truly being satisfied in the continuum limit using convergence tests.

We avoid the singularity of the space-time at $r=0$ excising a chunk of the domain inside the event horizon \cite{excision}. Thus we define the radial domain such that $r\in[r_{in},r_{ext}]$ with $0<r_{in}<r_{horizon}$ and $r_{ext}$ as big as possible. Provided we use Eddington-Finkesltein slices to describe the black hole space-time, light cones at $r_{in}$ are open and there is no need to impose any boundary conditions there. On the other hand, at $r=r_{ext}$ we apply outgoing radiative boundary conditions \cite{Jonathan}. Even though, in all our runs we make sure the exterior boundary is causally disconnected from the region where we measure physical quantities, in order to avoid the pollution from numerical errors reflected from the exterior boundary.

We update the gauge functions $\alpha$ and $\beta^r$ using the conditions $\alpha/\sqrt{\gamma_{rr}} + \beta^r = 1$ and $-2\alpha K_{\theta\theta} + \beta^r \partial_r \gamma{\theta\theta}=0$ that guarantee that $t+r$ is in ingoing null coordinate and that $\partial_t \gamma_{\theta\theta}=0$ respectively \cite{Jonathan}.

\subsection{Diagnostics}

Since we want to study the accreted scalar field, it is useful to track the apparent horizon of the black hole, which provides an approximate growth rate of the black hole mass, which further indicates the correct accreted scalar field. We achieve this by locating the outermost marginally trapped surface, which obeys the condition 

\begin{equation}
\Theta = \frac{\partial_r \gamma{\theta\theta}}{\gamma_{\theta\theta}\sqrt{\gamma_{rr}}}
-\frac{2K_{\theta\theta}}{\gamma_{\theta\theta}} =0, \label{eq:ah}
\end{equation}

\noindent which is a procedure we can practice on the fly during the evolution \cite{Jonathan}. Then we define the radius at which this happens as the apparent horizon radius $r_{AH}=\sqrt{\gamma_{\theta\theta}{}_{AH}}$ in terms of the areal radius. Then we estimate the apparent horizon mass using $M_{AH}= r_{AH}/2$. 
Since our results converge with second order due to the discretization along the spatial direction, the final apparent horizon mass is a Richardson extrapolation of our results using the two finest resolutions in all our calculations.

We also locate the event horizon because it is a gauge independent 3-surface unlike the apparent horizon which is a gauge dependent 2-sruface. The reason to locate the event horizon is that we want to make sure the event horizon lies outside the apparent horizon and close to it after the accretion of the scalar field has finished. We locate the event horizon as the attractor surface of inward null rays triggered from the future and from outside the black hole.

In order to make sure that the horizon growth is consistent with the mass of the space-time including the contribution of the scalar field, we measure the Misner-Sharp mass \cite{mtw} defined for our space-time as

\begin{equation}
M_{MS} = \frac{1}{2}\sqrt{\gamma_{\theta\theta}} \left( 1 - \frac{1}{4} \frac{(\partial_r \gamma_{\theta\theta})^2}{\gamma_{rr}\gamma{\theta\theta}} + \frac{K_{\theta\theta}^{2}}{\gamma_{\theta\theta}}\right)
\end{equation}

\noindent which allows us to estimate the ADM mass $M_{ADM}=\lim_{r\rightarrow \infty} M_{MS}$. In this way we can diagnose whether or not the final mass of the black hole's horizon corresponds to $M_{ADM}$.

Finally we also track the constraints (\ref{eq:Mom}) and practice the required convergence tests they must satisfy. All our results assume the radial and time coordinate are written in units of $M$, where $M$ is the initial mass of the black hole's event horizon.

\subsection{Example}

\begin{figure}
\includegraphics[width=4.25cm]{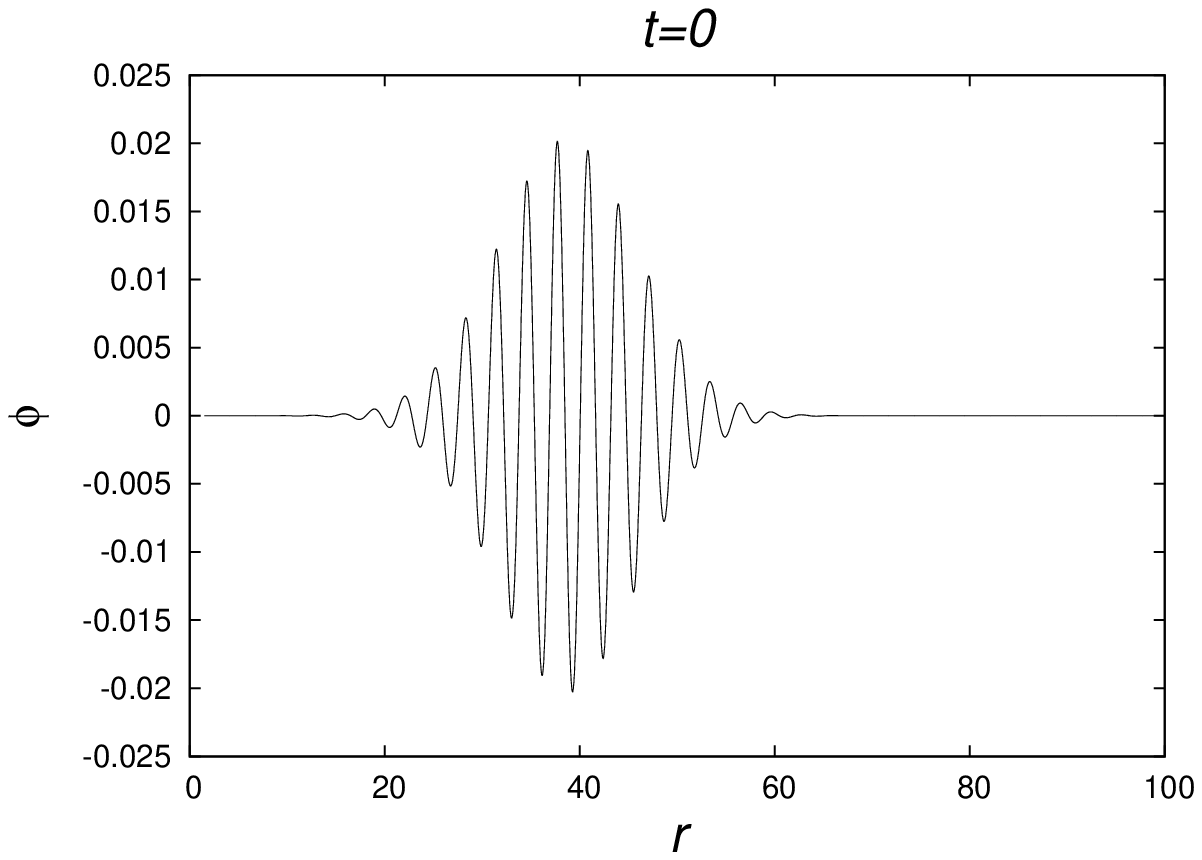}
\includegraphics[width=4.25cm]{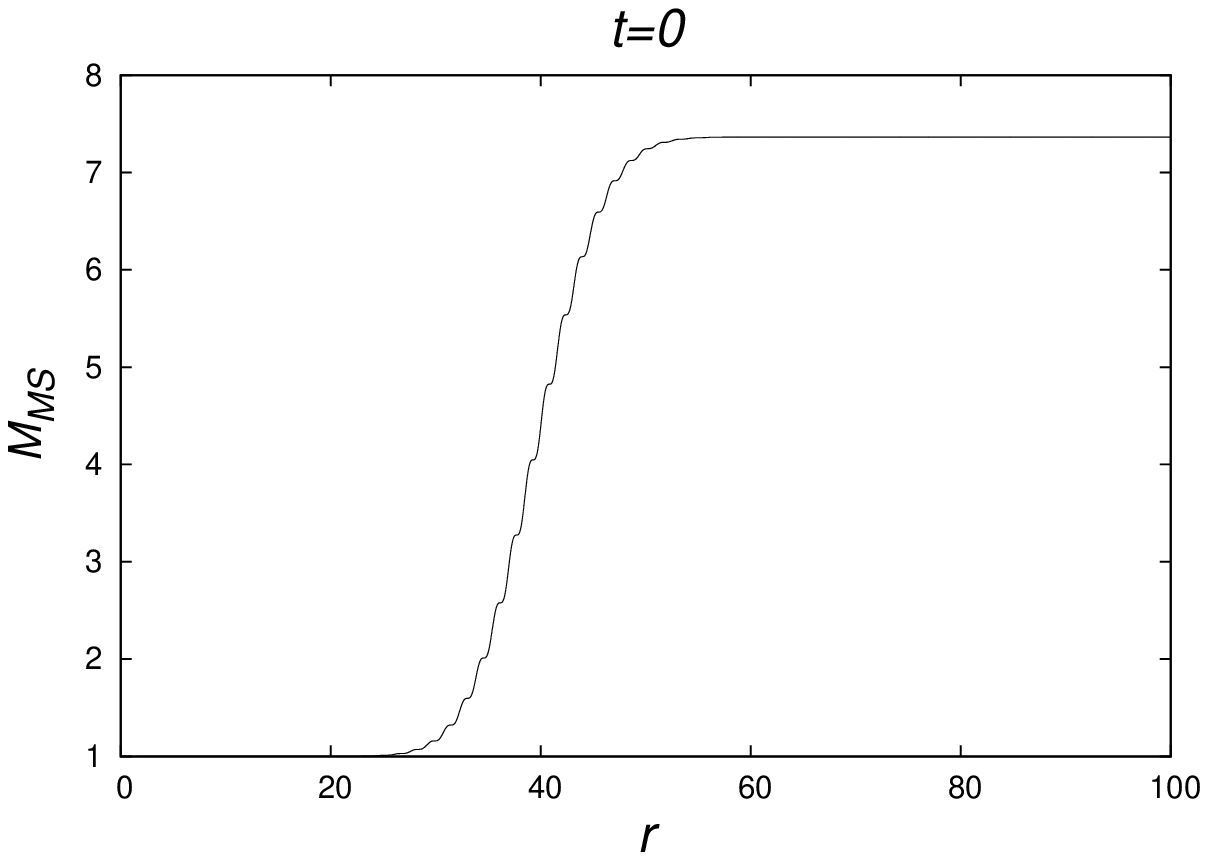}
\includegraphics[width=4.25cm]{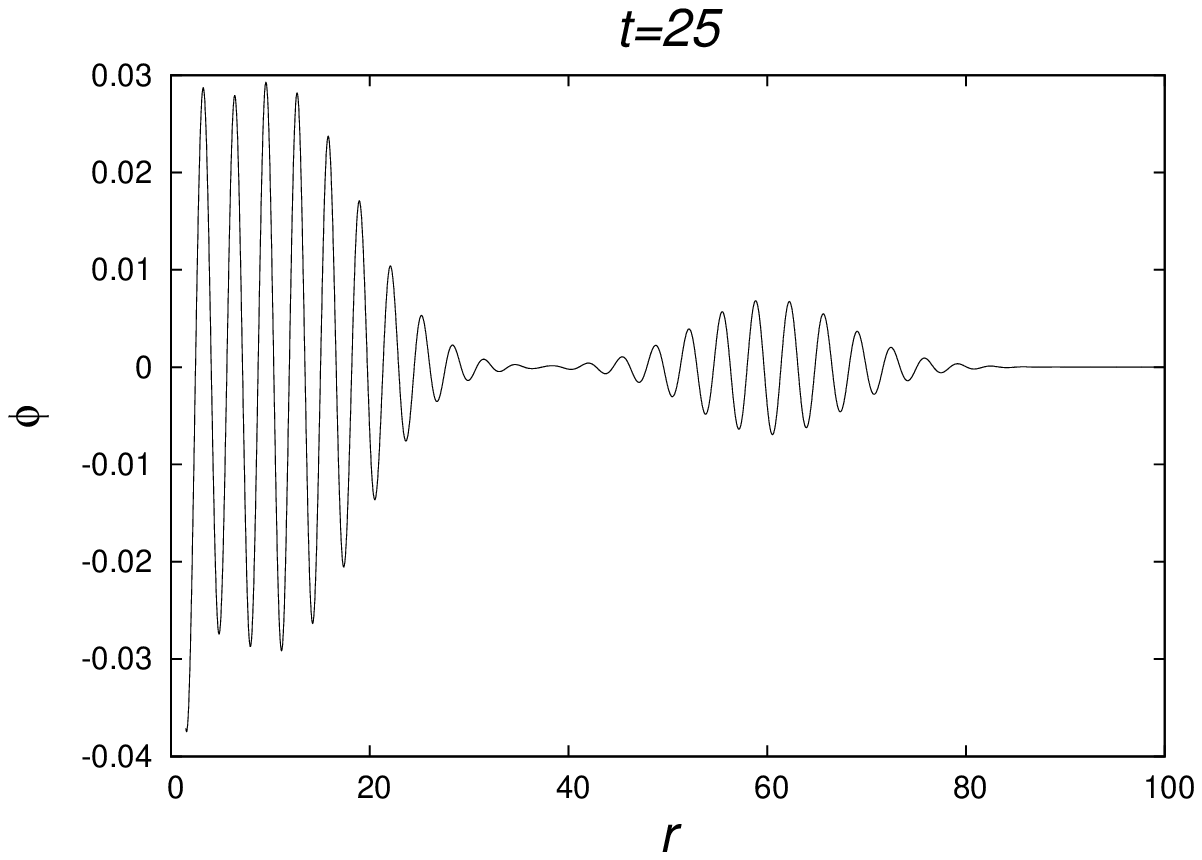}
\includegraphics[width=4.25cm]{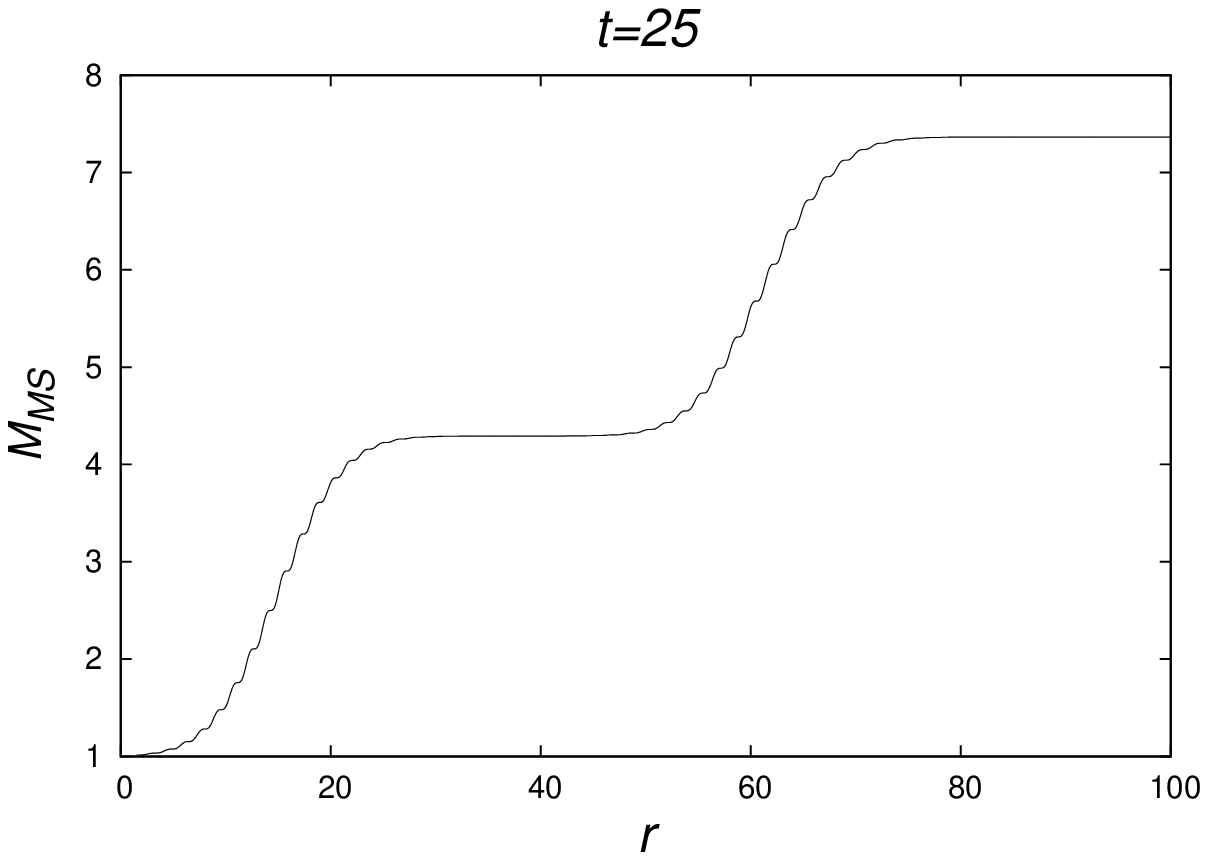}
\includegraphics[width=4.25cm]{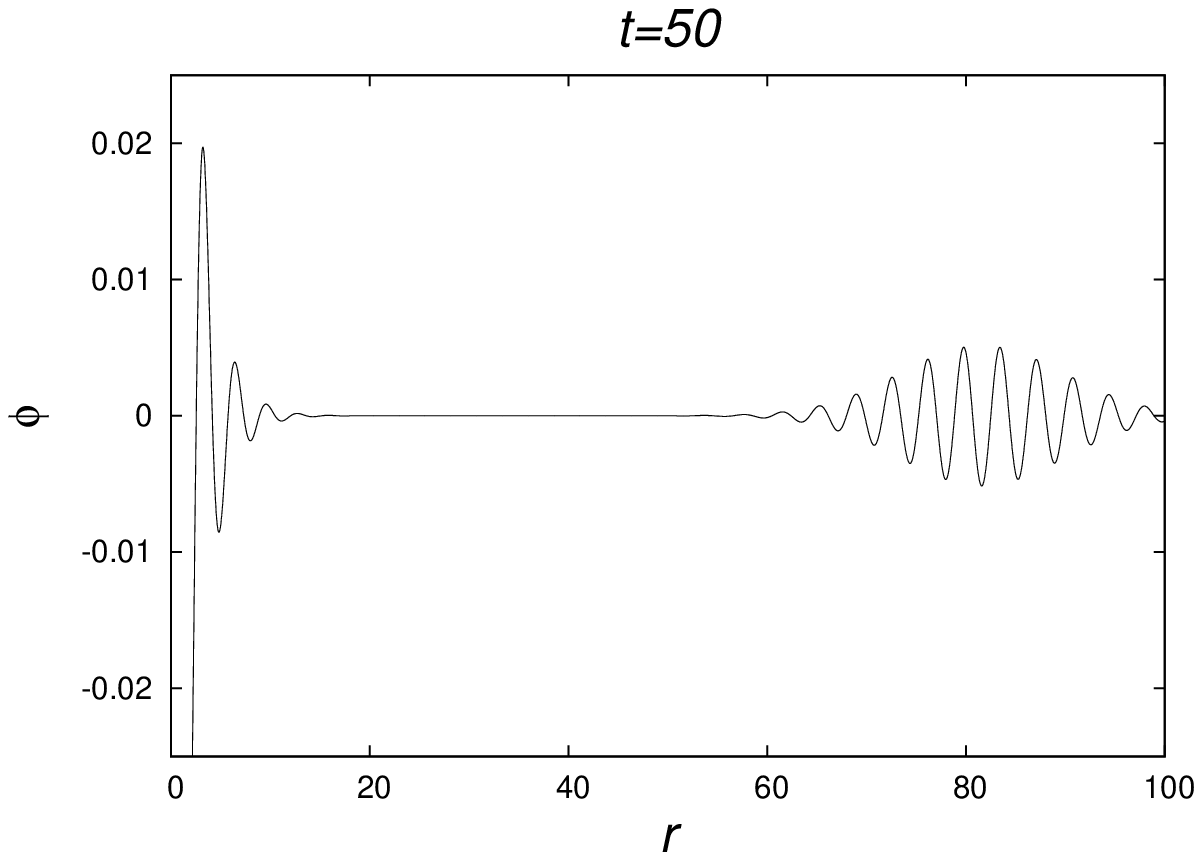}
\includegraphics[width=4.25cm]{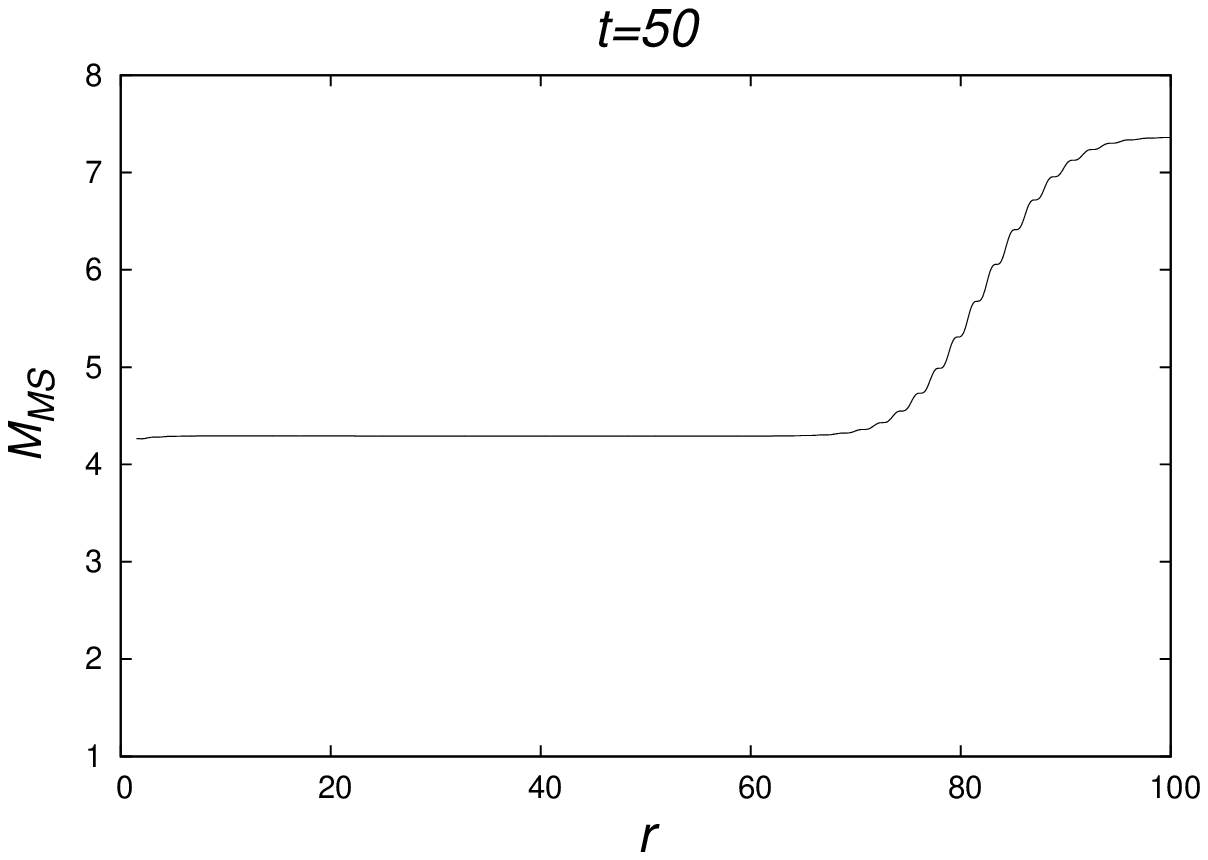}
\includegraphics[width=4.25cm]{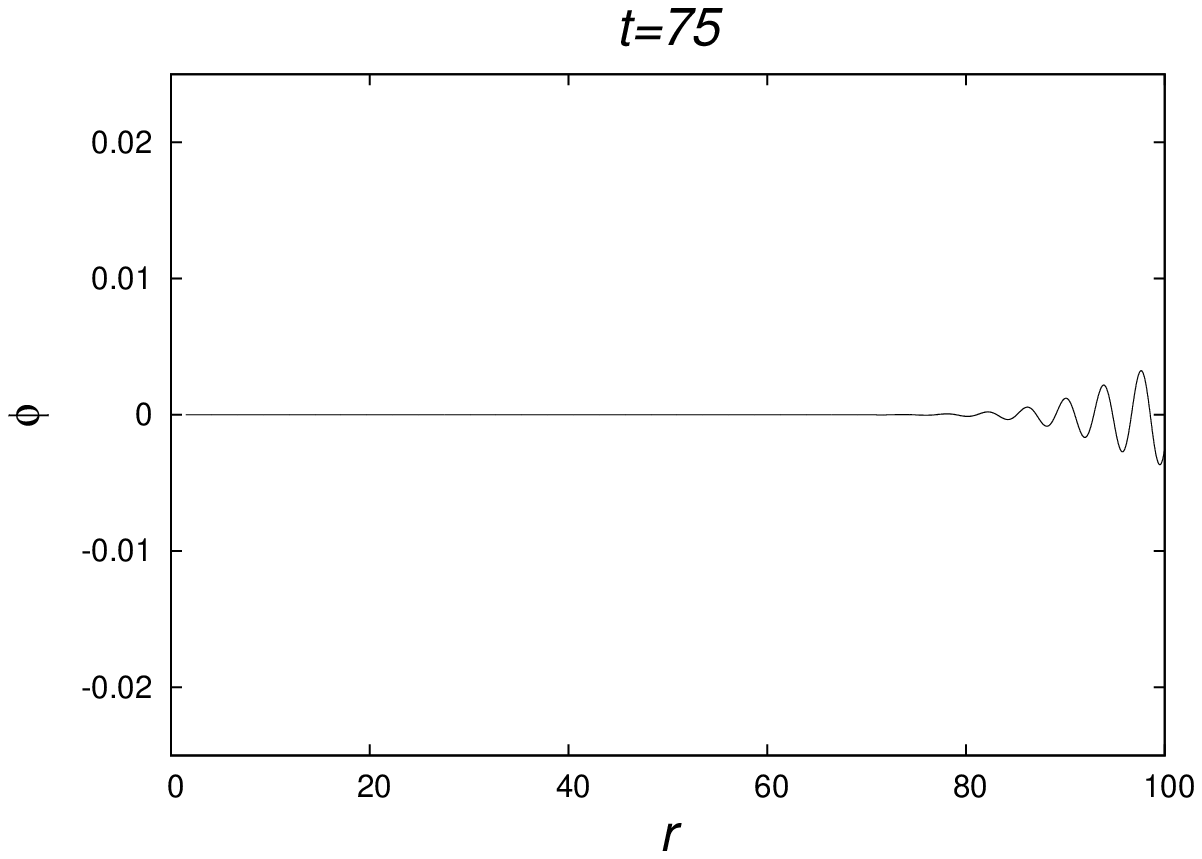}
\includegraphics[width=4.25cm]{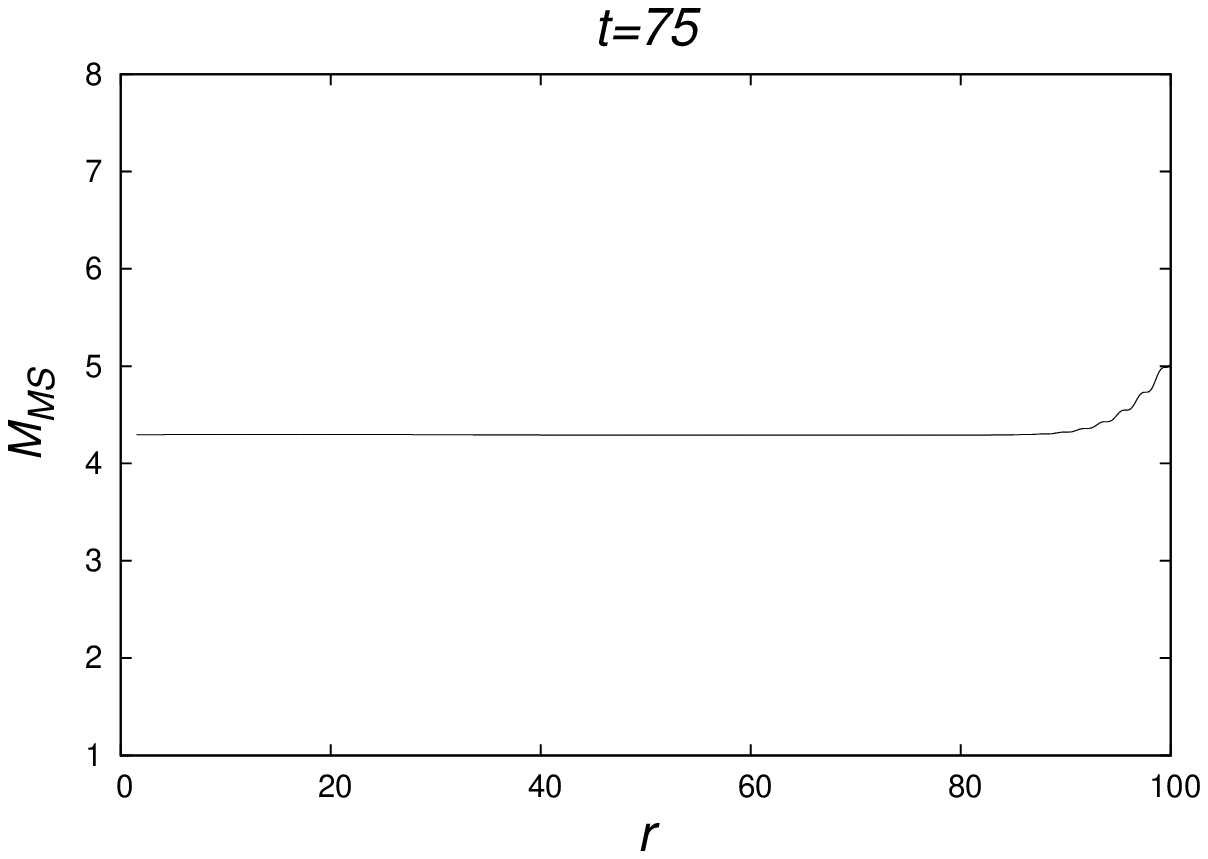}
\includegraphics[width=7.5cm]{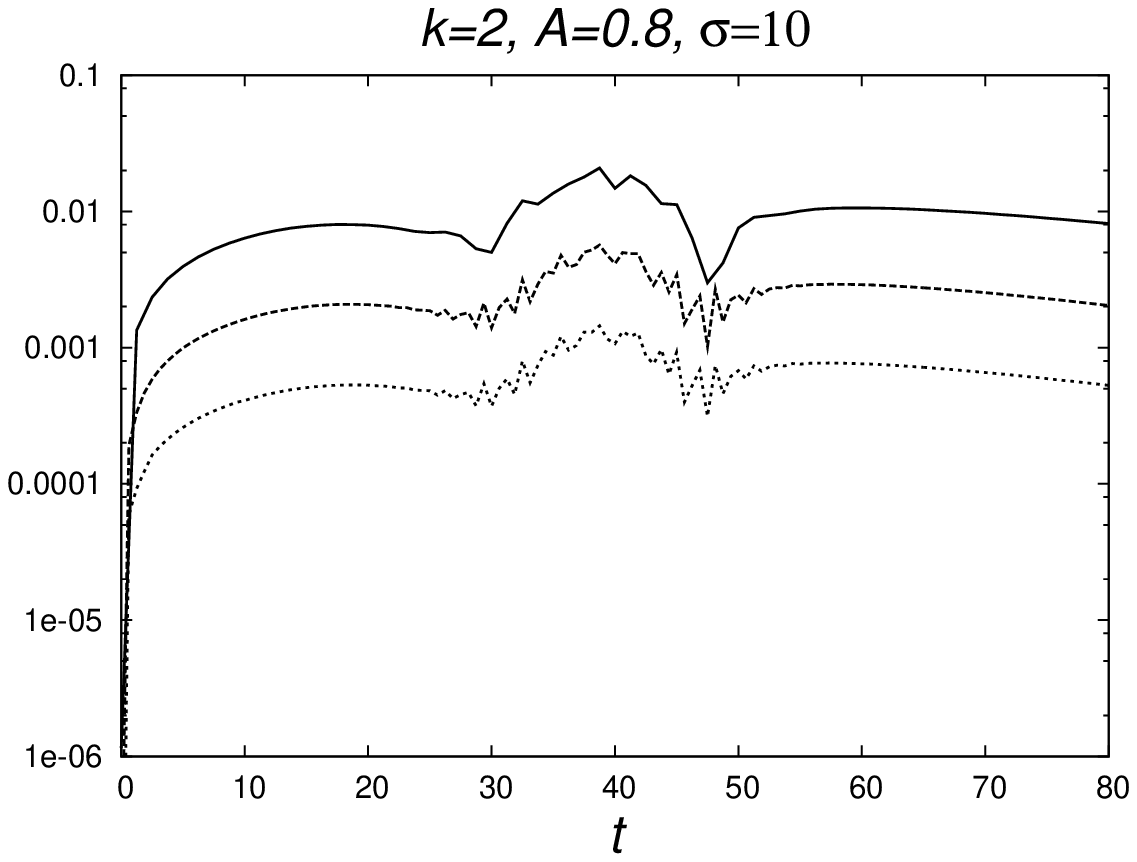}
\caption{\label{fig:example} We show one of the cases of our sample in detail, the one corresponding to
$k=2$, $A=0.8$ and $\sigma=10$. 
We show snapshots of the pulse moving toward the black hole. As mentioned in the text, the initial pulse
splits into two, one moving inward and the other outward. 
We show on the right sides the Misner-Sharp mass for the same snapshots and show that it approaches 
a constant value that includes only the incoming pulse contribution. The Misner-Sharp mass becomes a constant after the two pulses carry their respective mass-energy contributions. This constant value 
is what we call our $M_{ADM}$ and is free of the contribution of the outgoing pulse. Finally we present as a sample of all our convergence tests the $L_2$ norm of the Hamiltonian 
constraint. The resolutions used in our runs are $\Delta r_1 = 0.0125M$ 
(continuous line), $\Delta r_2 = \Delta r_1 /2$ (dashed line) and 
$\Delta r_3 = \Delta r_2 /2$ (dotted line).}
\end{figure}

Now we present in detail one of our simulations. In Fig. \ref{fig:example} we illustrate how we estimate the mass of the space-time. Some snapshots are shown for the scalar field and for the Misner-Sharp mass. We show on the left panels how the initial pulse splits into two pulses, one moving toward the black hole and the other moving outwards. On the right panels we show the Misner-Sharp mass and illustrate how the outgoing pulse carries a given amount of the space-time mass. The mass measured once the system has stabilized is what we call the ADM mass $M_{ADM}$, around $t \sim 75$. This would be the mass of a space-time if it would only containing the ingoing pulse. Finally in Fig. \ref{fig:example} we present the convergence test of the $L_2$ norm of the Hamiltonian constraint in order to illustrate how all our runs behave.

\begin{figure}
\includegraphics[width=8cm]{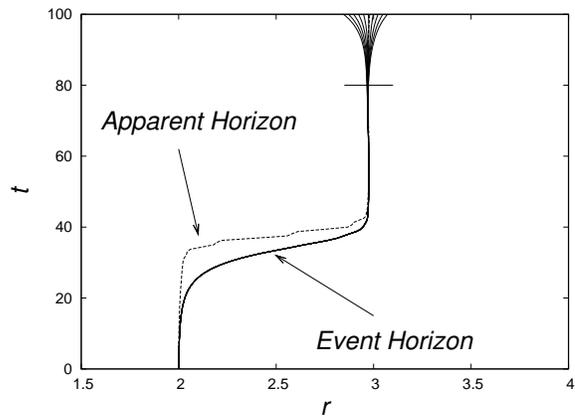}
\caption{\label{fig:ehorizon} We show one of the cases of our sample in detail, the one corresponding to
$k=1$, $A=0.8$ and $\sigma=5$. This plot shows how the apparent horizon and event horizon coincide both at initial time and after the black hole has stabilized after the accretion of the incoming scalar field pulse. We also show a set of outgoing null rays, or equivalently ingoing past directed null rays, that converge to a surface that happens to be the event horizon. The event horizon is only tracked up to $t\sim 80$, after which our null rays diverge.}
\end{figure}

In Fig. \ref{fig:ehorizon} we show other of our simulations showing both the apparent and the event horizons. What we want to illustrate is that on the one hand the apparent horizon lies always inside the event horizon, which is consistent with the energy conditions of the space-time and also, that once the black hole has accreted the incoming scalar field, both horizons coincide, which authorizes us to use the apparent horizon as the surface to monitor the mass of the black hole in time.

\section{Results}
\label{sec:results}

We split the parameter space in terms of the wave number $k$, which is one of the properties of the propagation speed of the wave packet. In fact it was found for a fixed background that for $k<1$ the absorption rates diminish. Thus we choose four values of $k=0,~0.5,~1,~2$, for which we choose a rather spread set of parameter values $A= 0.5,~0.8$ and $\sigma = 1,~5,~10$. With these parameters we sample different scalar field contributions to the energy density of the space-time, length scales and number of nodes of the wave function. In \cite{UrenaFernandez} the parameter $\sigma$ is related to the variance on $k$ provided the wave number lies around $k_0$ such that $\langle (k-k_0)^2 \rangle = 1/\sigma^2$ and full absorption was found for $k > 1$, and only partial absorption in other cases. We choose two values of $k$ smaller than the threshold $k=1$ and two bigger in order to confirm or contradict the predictions in \cite{UrenaFernandez}. 

\begin{figure}[htp]
\includegraphics[width=4.25cm]{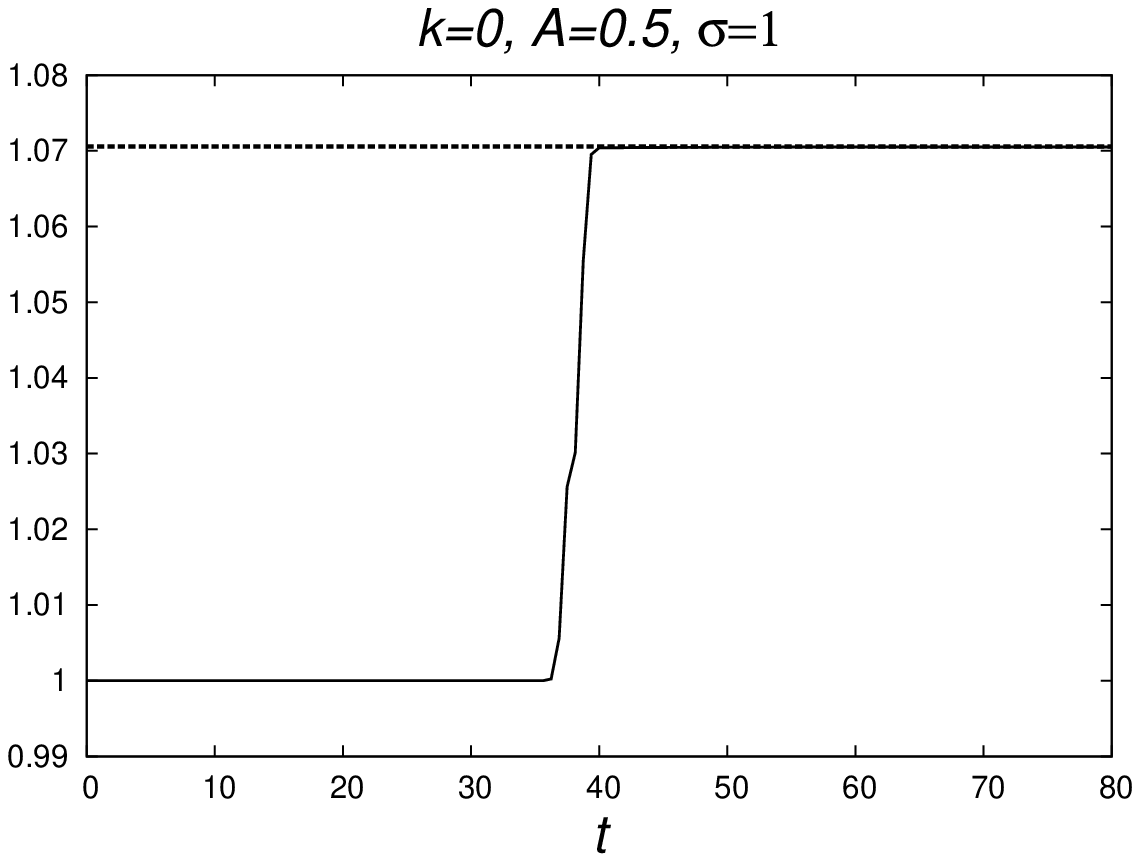}
\includegraphics[width=4.25cm]{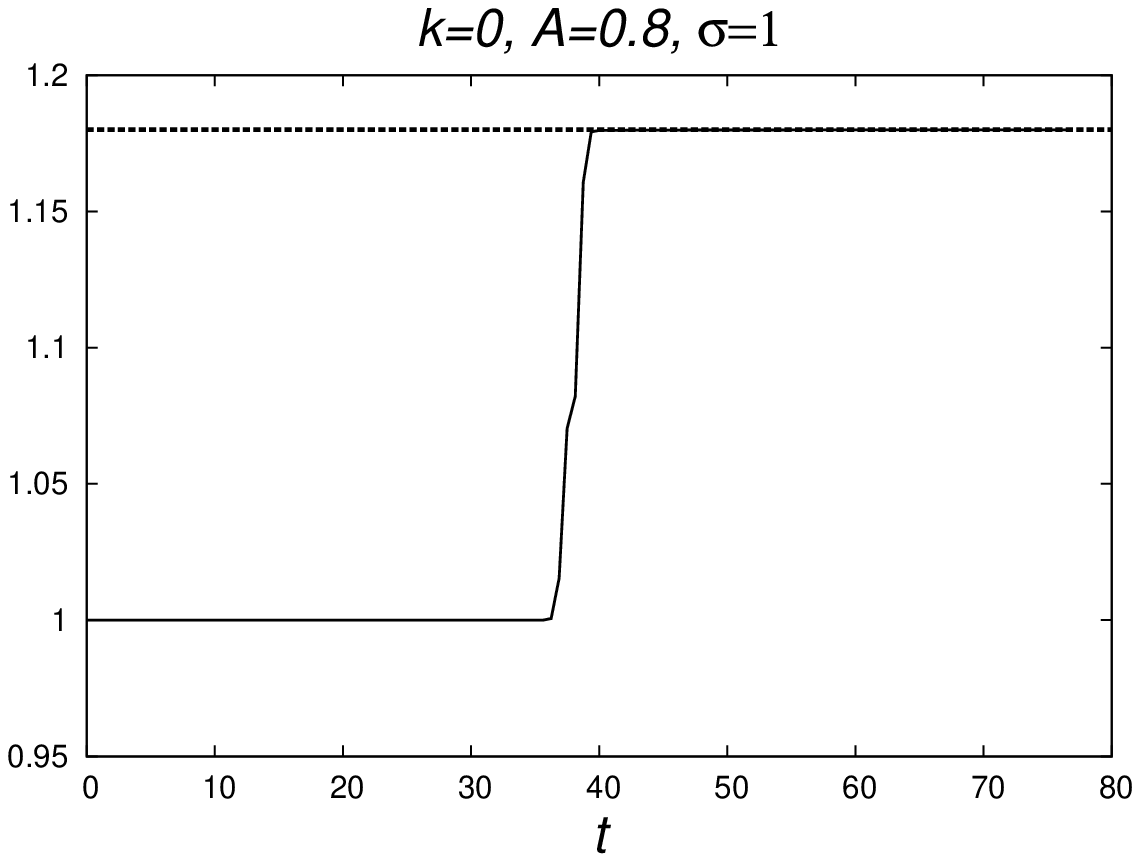}
\includegraphics[width=4.25cm]{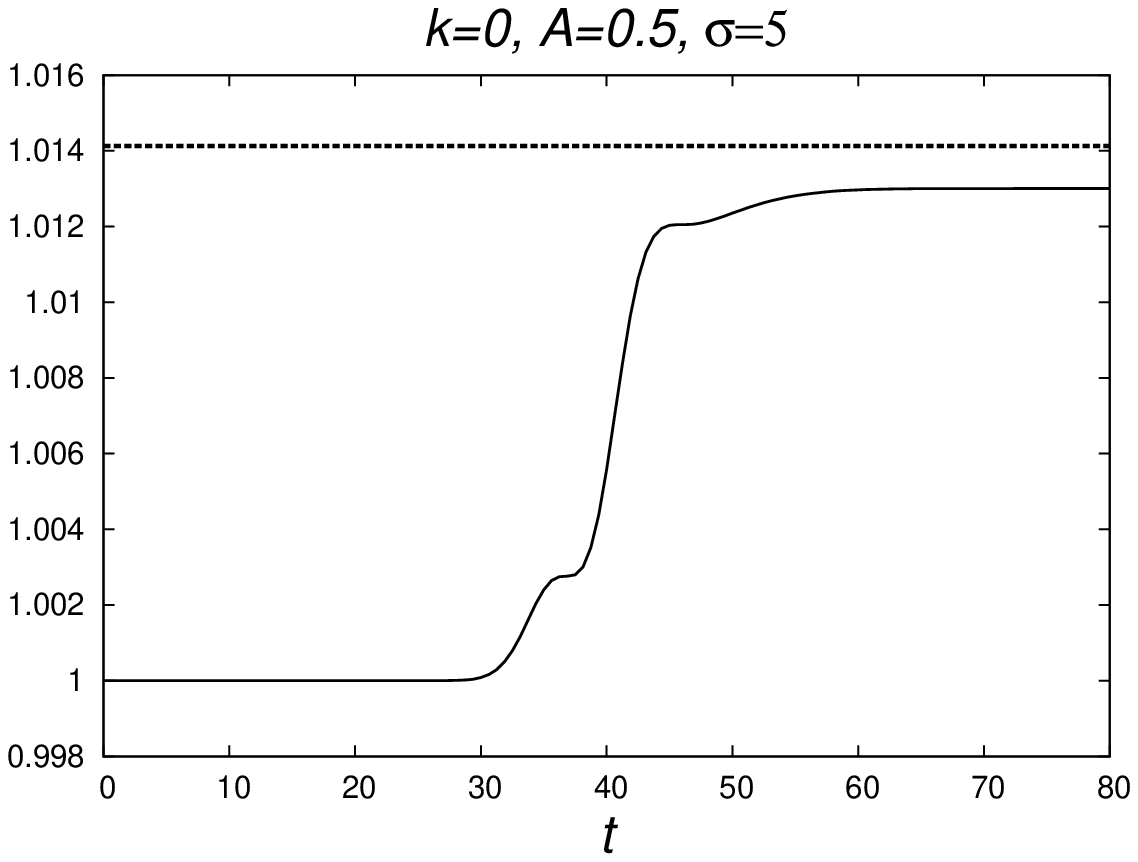}
\includegraphics[width=4.25cm]{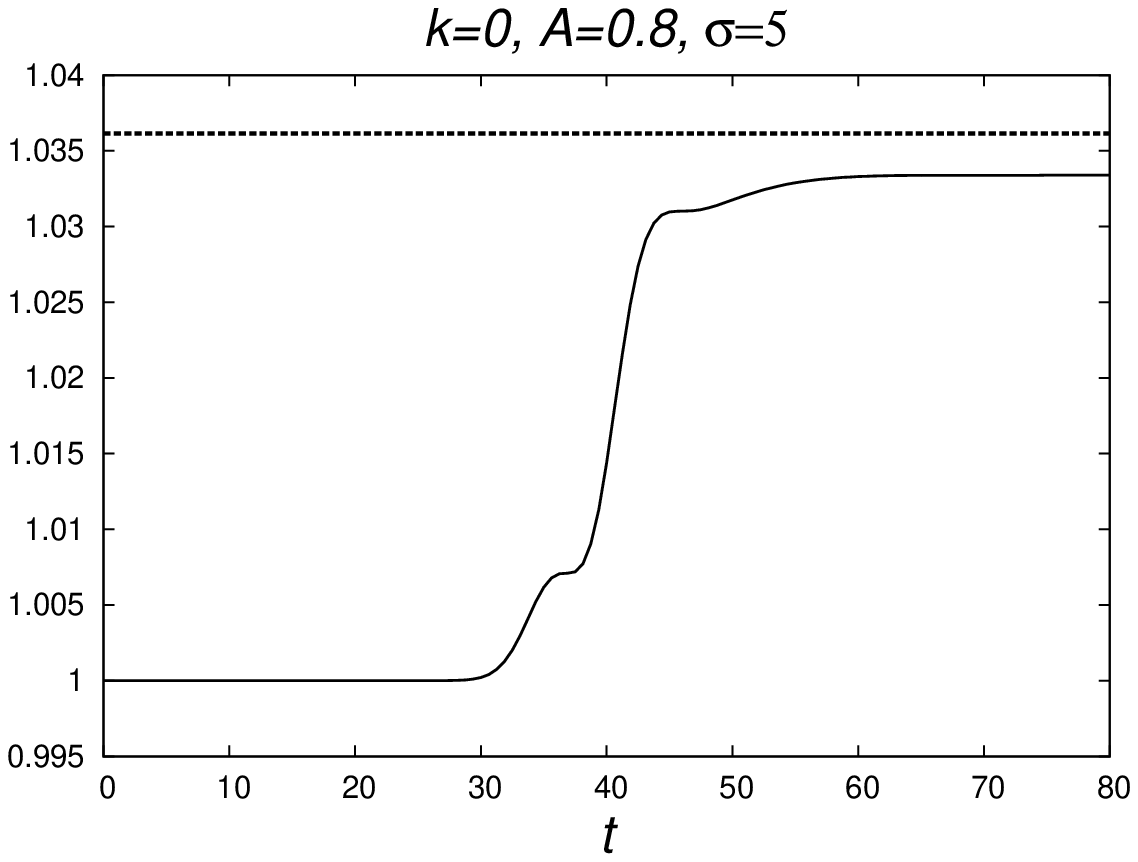}
\includegraphics[width=4.25cm]{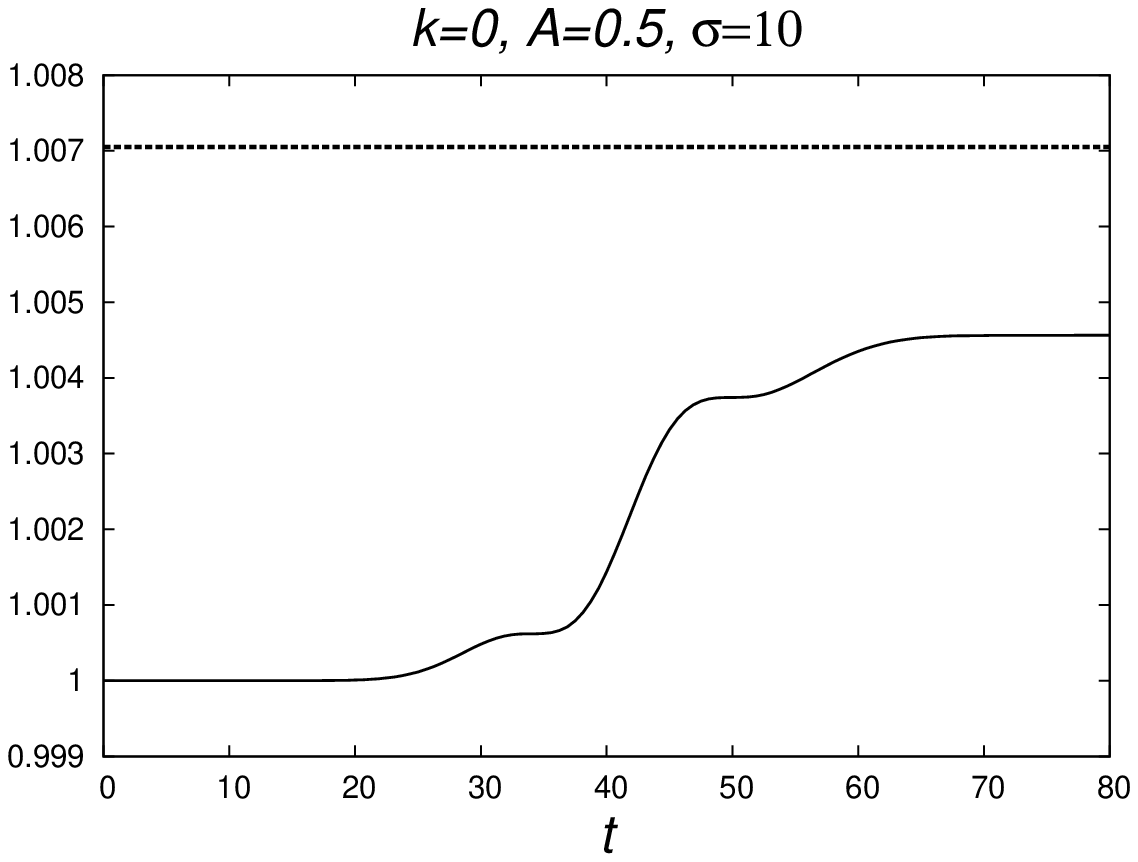}
\includegraphics[width=4.25cm]{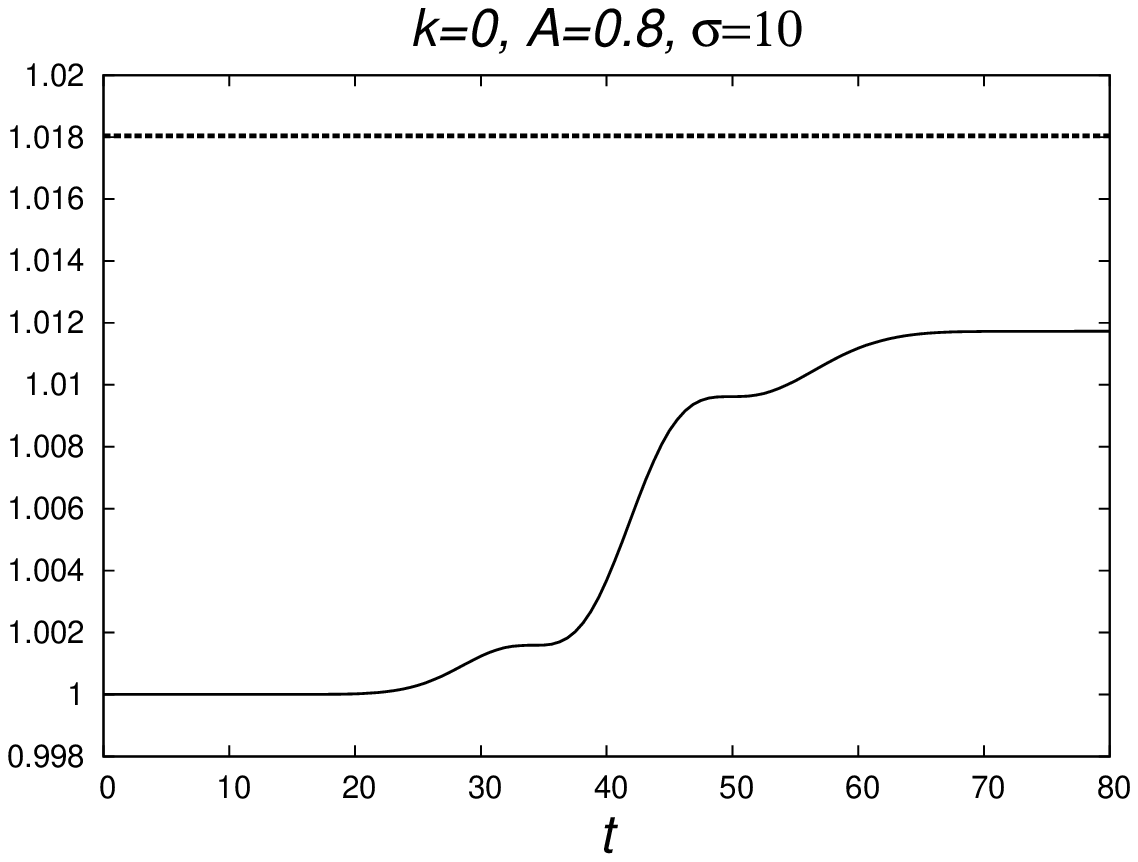}
\caption{\label{fig:k_0} We show the time evolution of the apparent horizon mass and the $M_{ADM}$ for the case $k=0$. The fact that the apparent horizon mass (continuous line) grows up to the value $M_{ADM}$ (dashed line) indicates that the whole incoming scalar field energy density has been absorbed, which happens only for $\sigma=1$, smaller than the Schwarzschild radius $r_S=2$, whereas for the bigger values $\sigma = 5, ~10$ a fraction of the scalar field is not being absorbed. The proportion of the field depends only on the value of $\sigma$, and we found that for $\sigma =5$ 92\% of the energy density of the scalar field has been absorbed and for $\sigma =10$ 67\%.}
\end{figure}

\begin{figure}[htp]
\includegraphics[width=4.25cm]{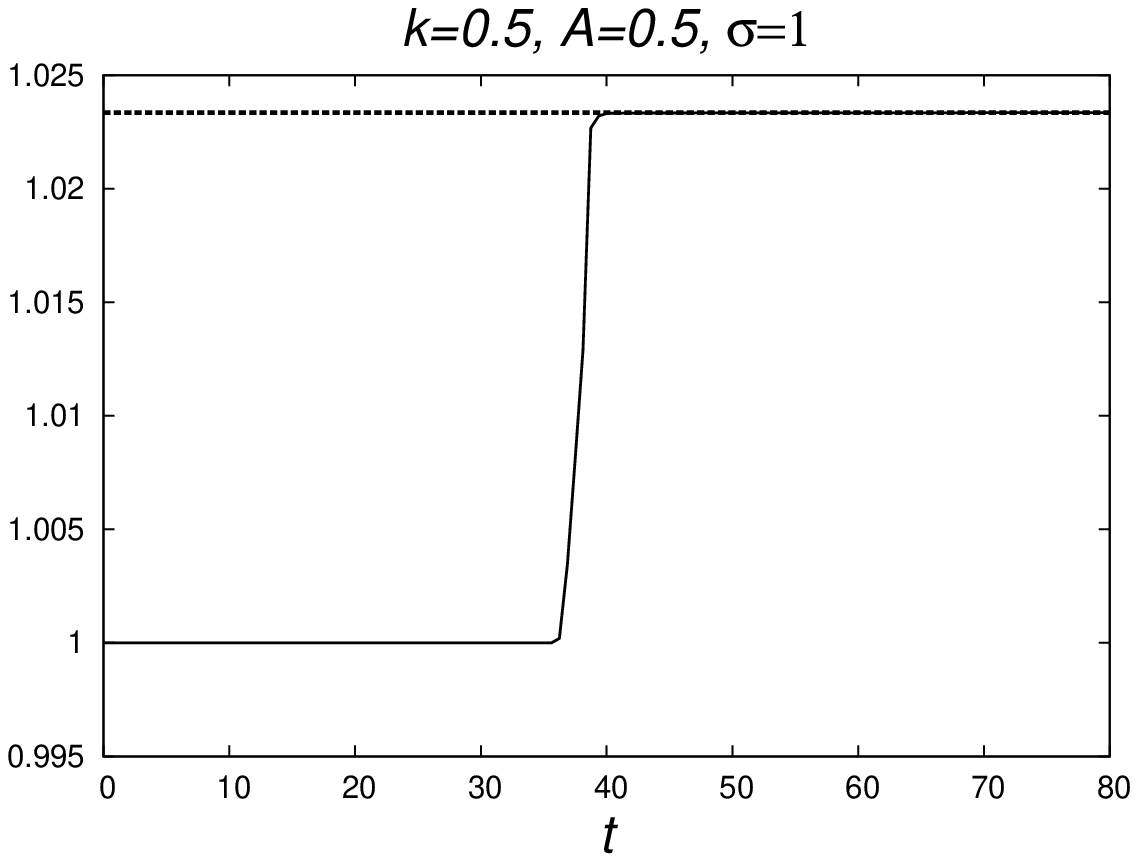}
\includegraphics[width=4.25cm]{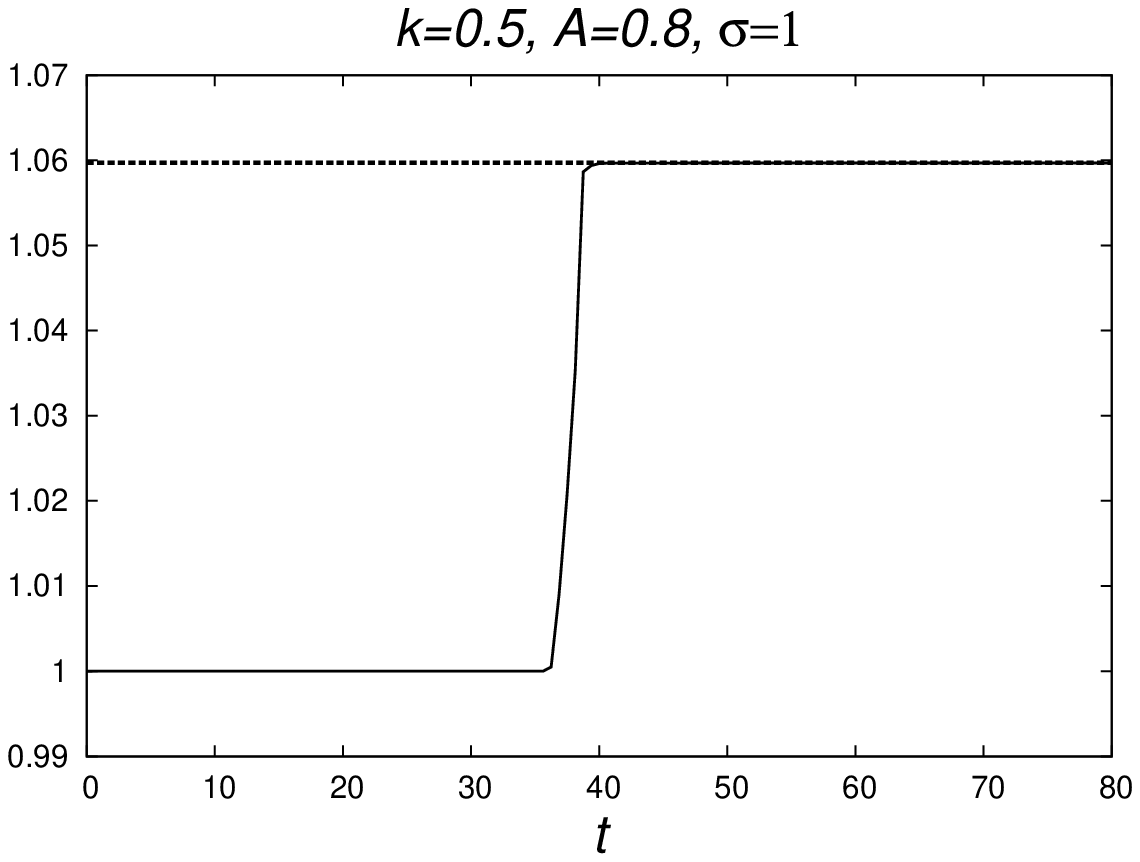}
\includegraphics[width=4.25cm]{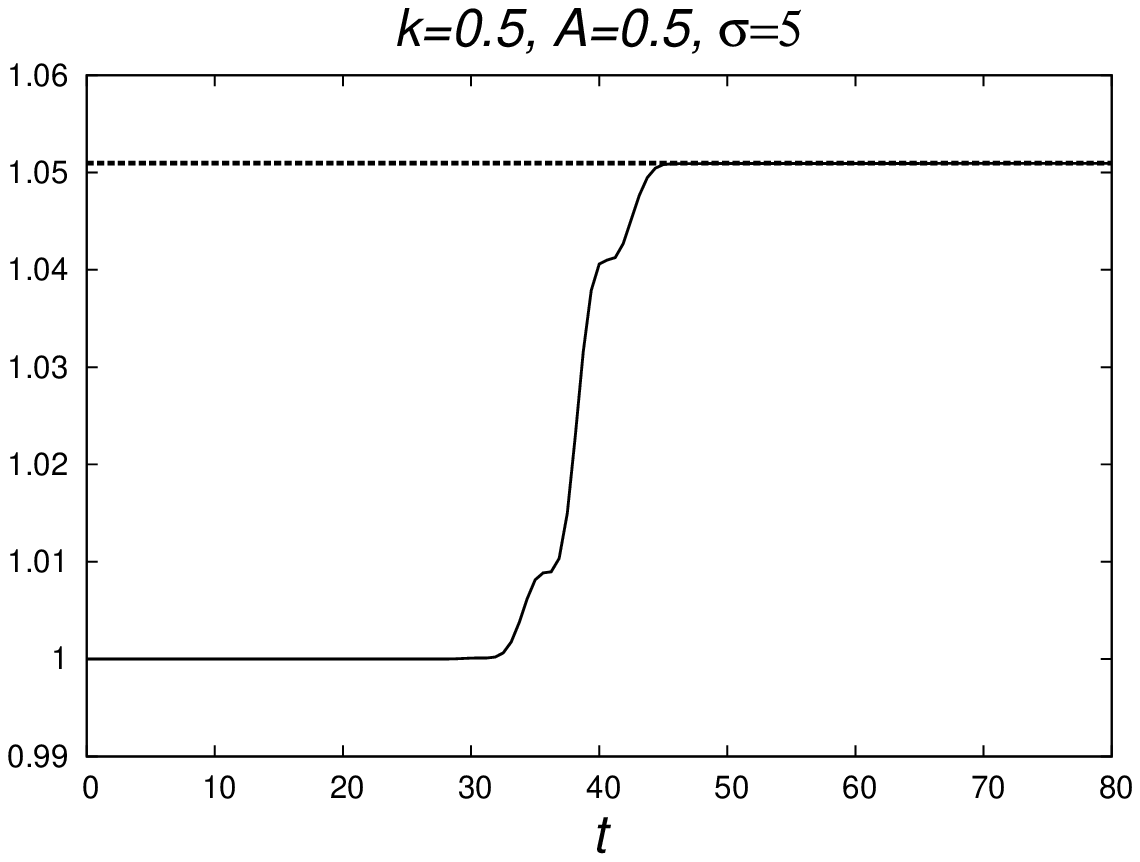}
\includegraphics[width=4.25cm]{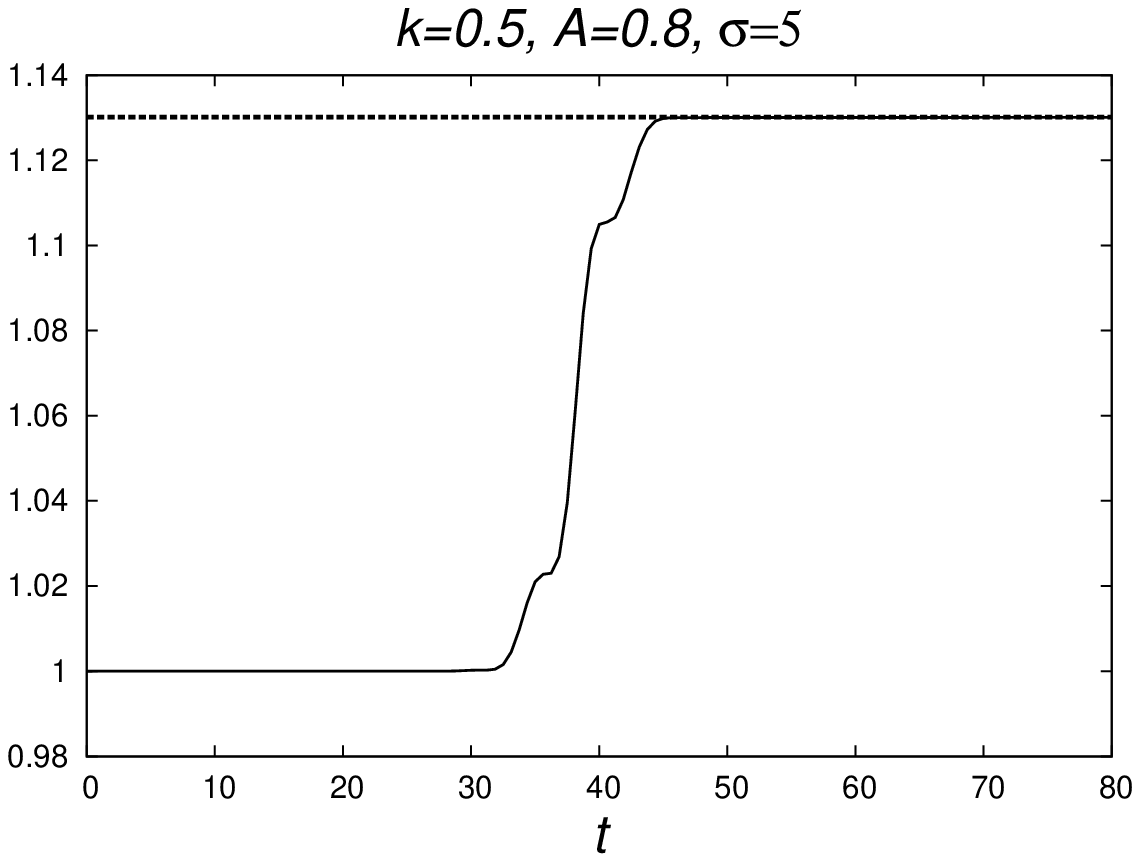}
\includegraphics[width=4.25cm]{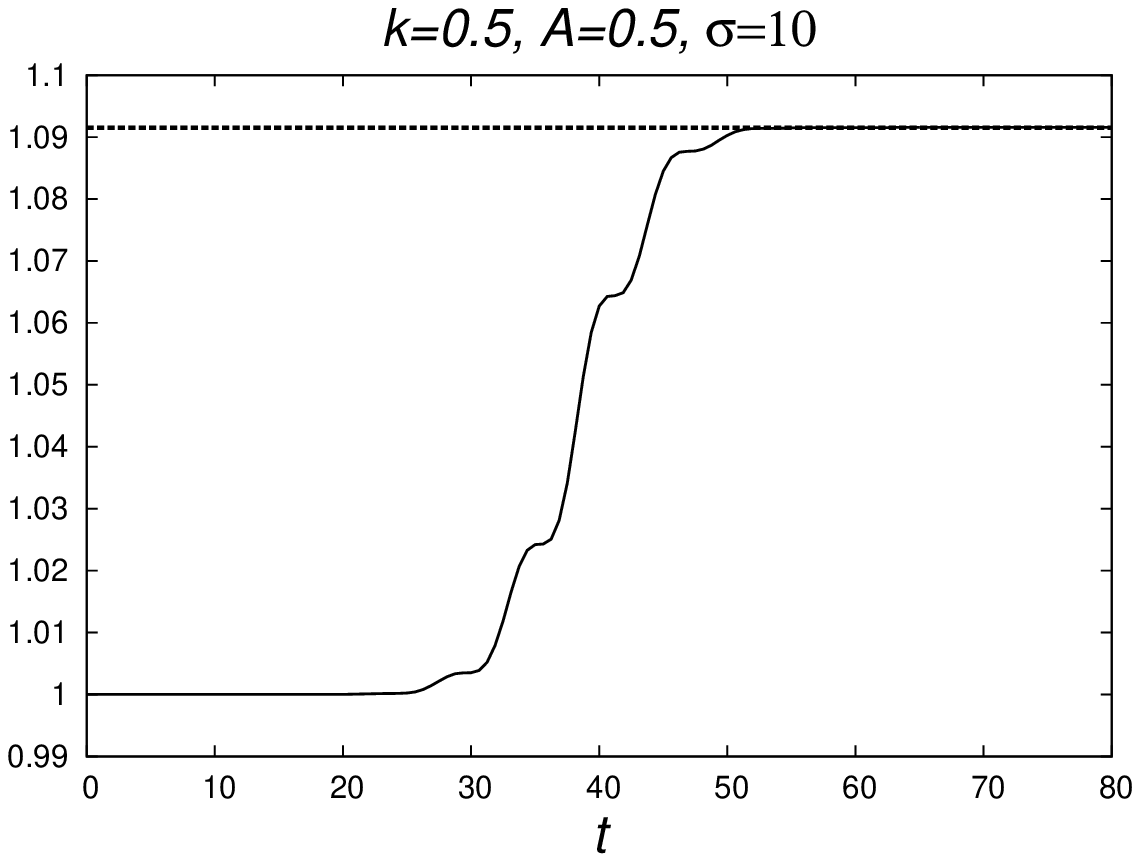}
\includegraphics[width=4.25cm]{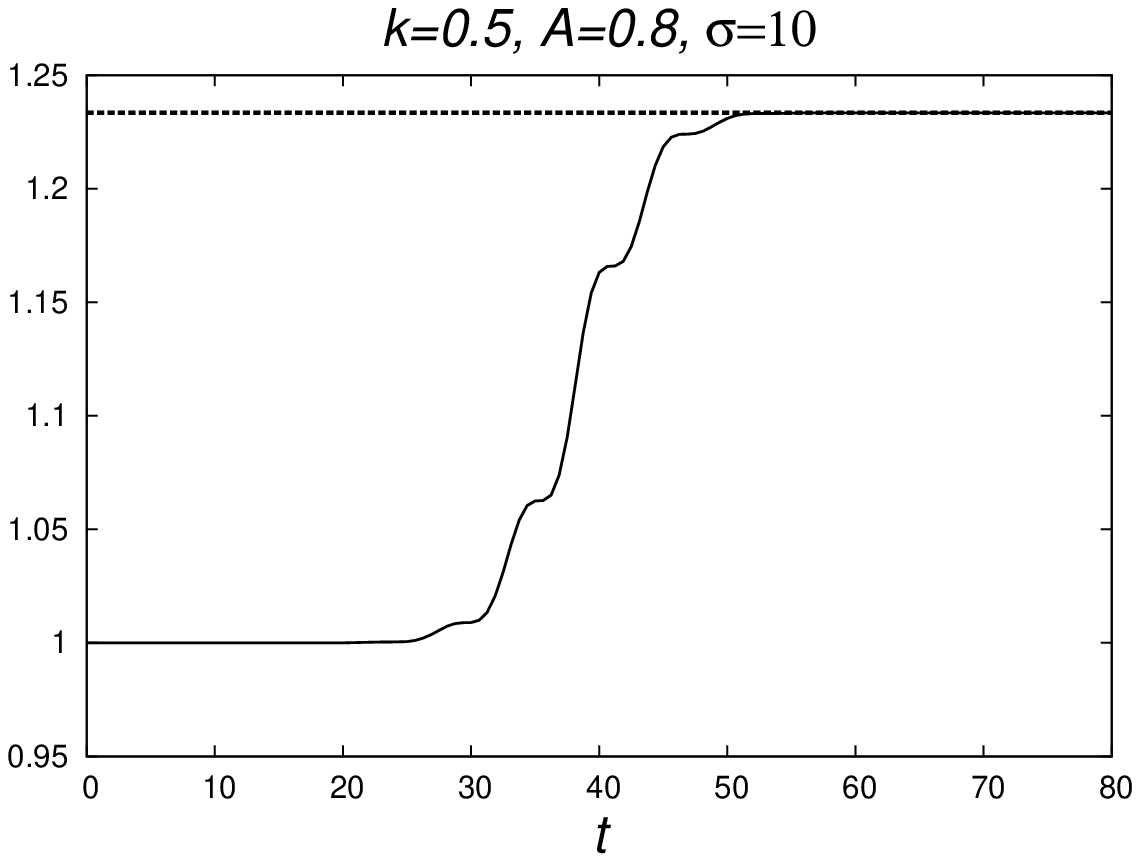}
\caption{\label{fig:k_0.5} We show the time evolution of the apparent horizon mass (continuous line) and the $M_{ADM}$ (dashed line) for the case $k=0.5$. Unlike the case $k=0$, in the present case full absorption was found in all cases.}
\end{figure}

\begin{figure}[htp]
\includegraphics[width=4.25cm]{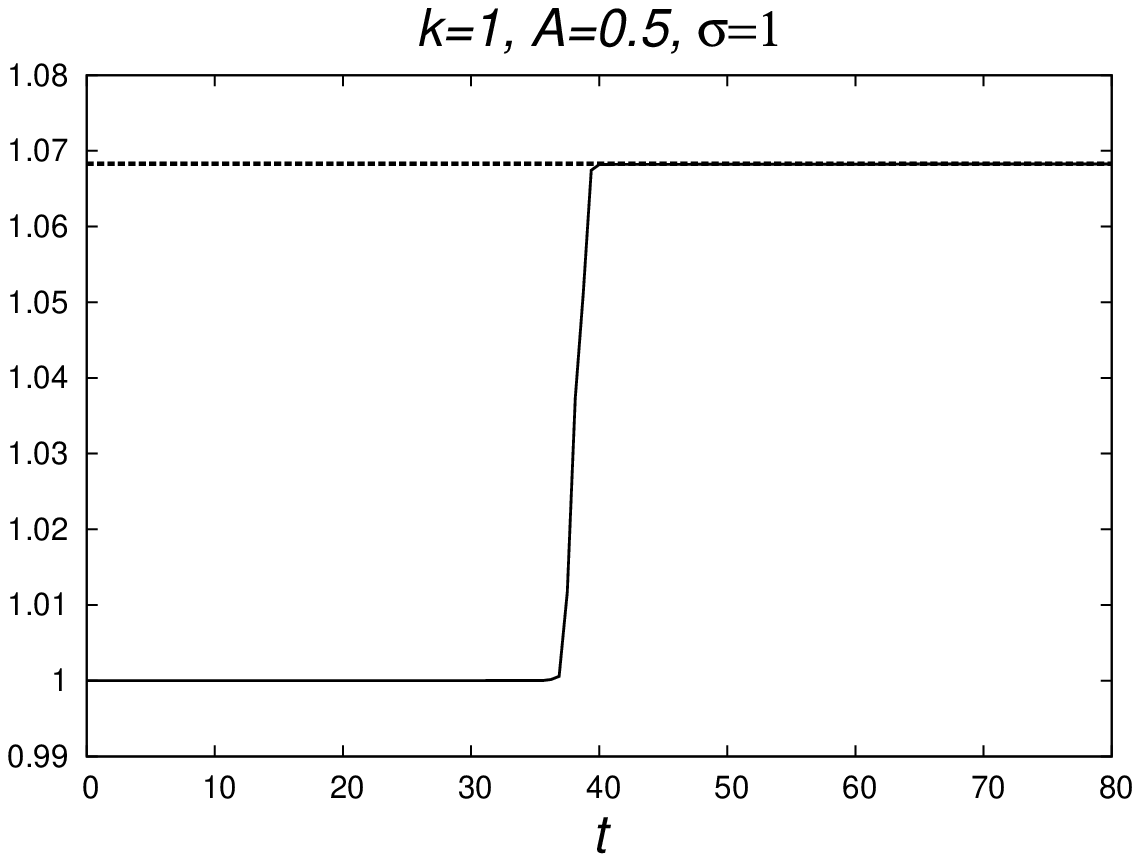}
\includegraphics[width=4.25cm]{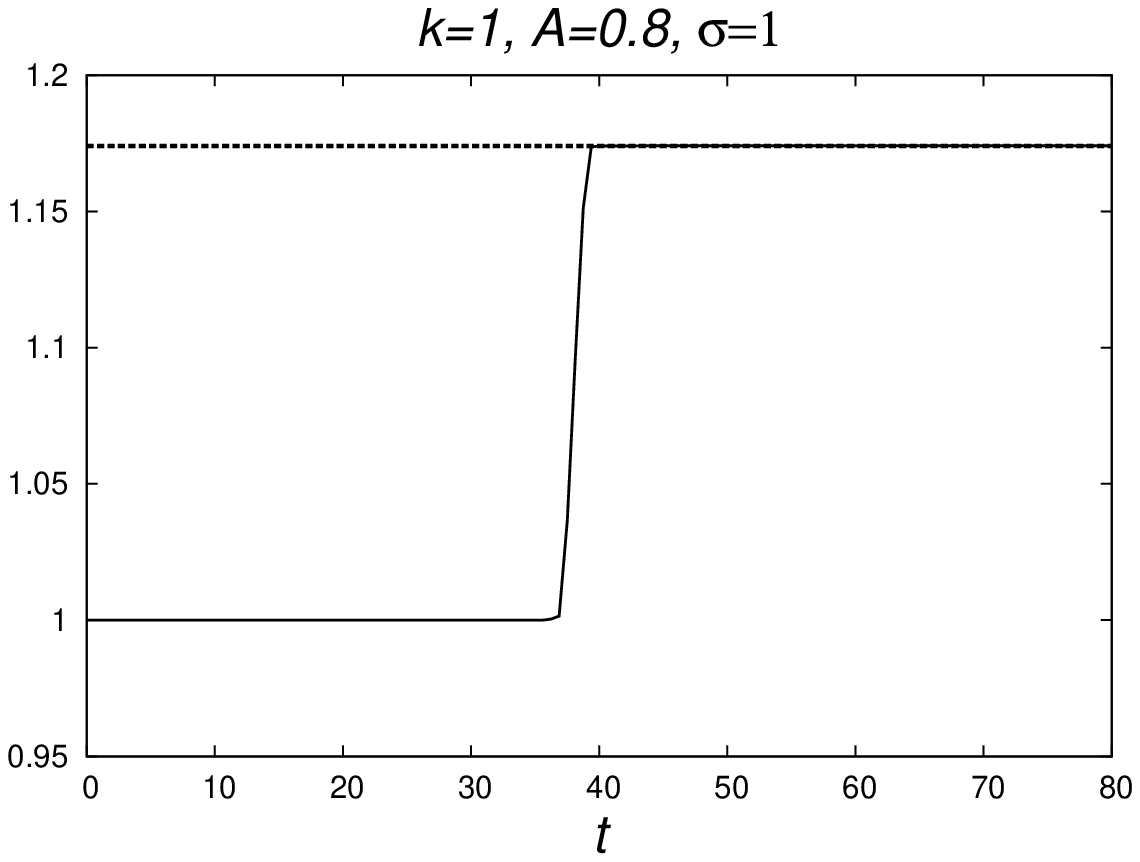}
\includegraphics[width=4.25cm]{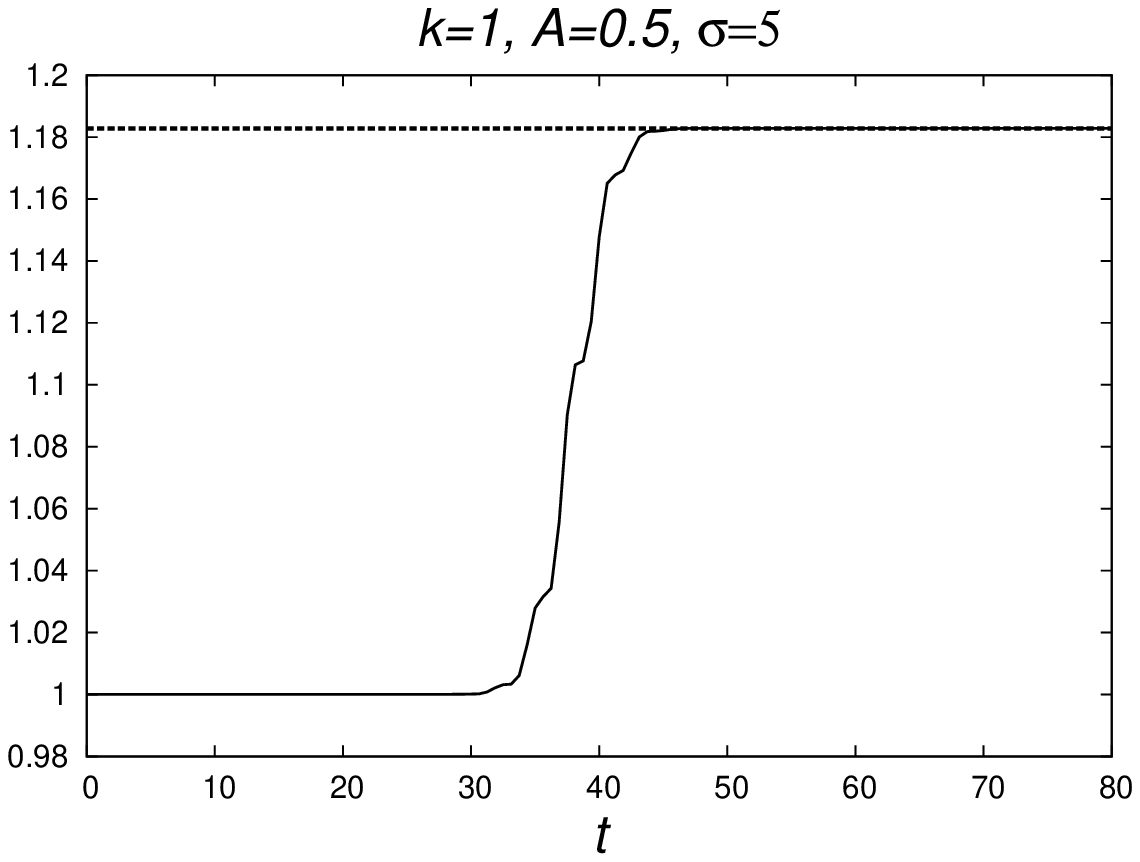}
\includegraphics[width=4.25cm]{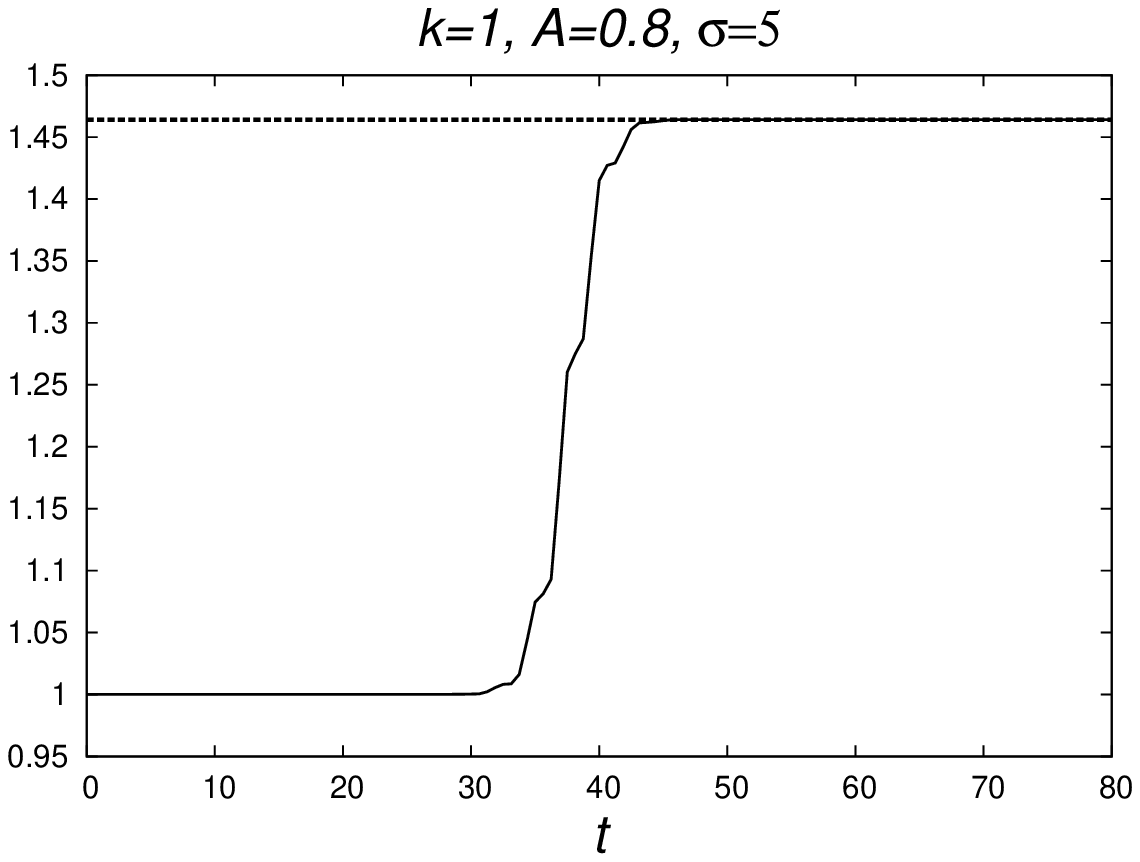}
\includegraphics[width=4.25cm]{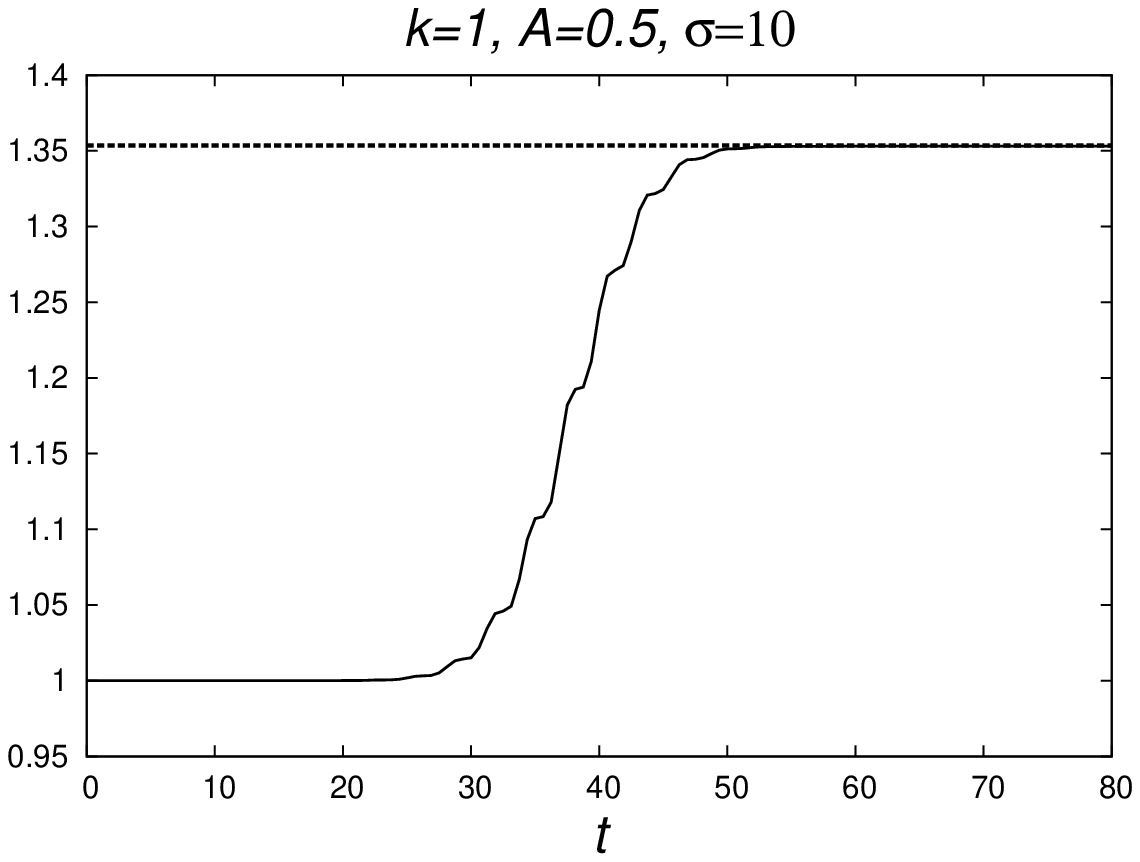}
\includegraphics[width=4.25cm]{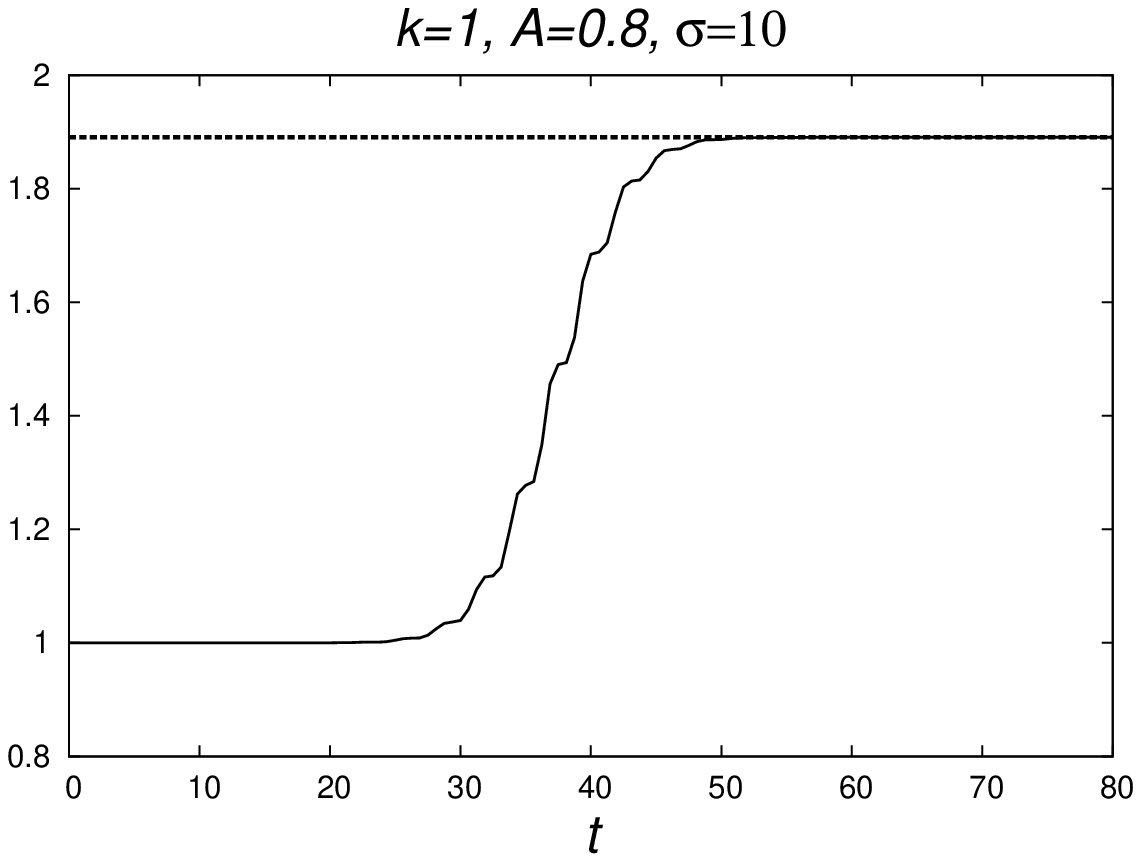}
\caption{\label{fig:k_1} We show the time evolution of the apparent horizon mass (continuous line) and the $M_{ADM}$ (dashed line) for the case $k=1$. Also full absorption was found in all cases.}
\end{figure}

\begin{figure}[htp]
\includegraphics[width=4.25cm]{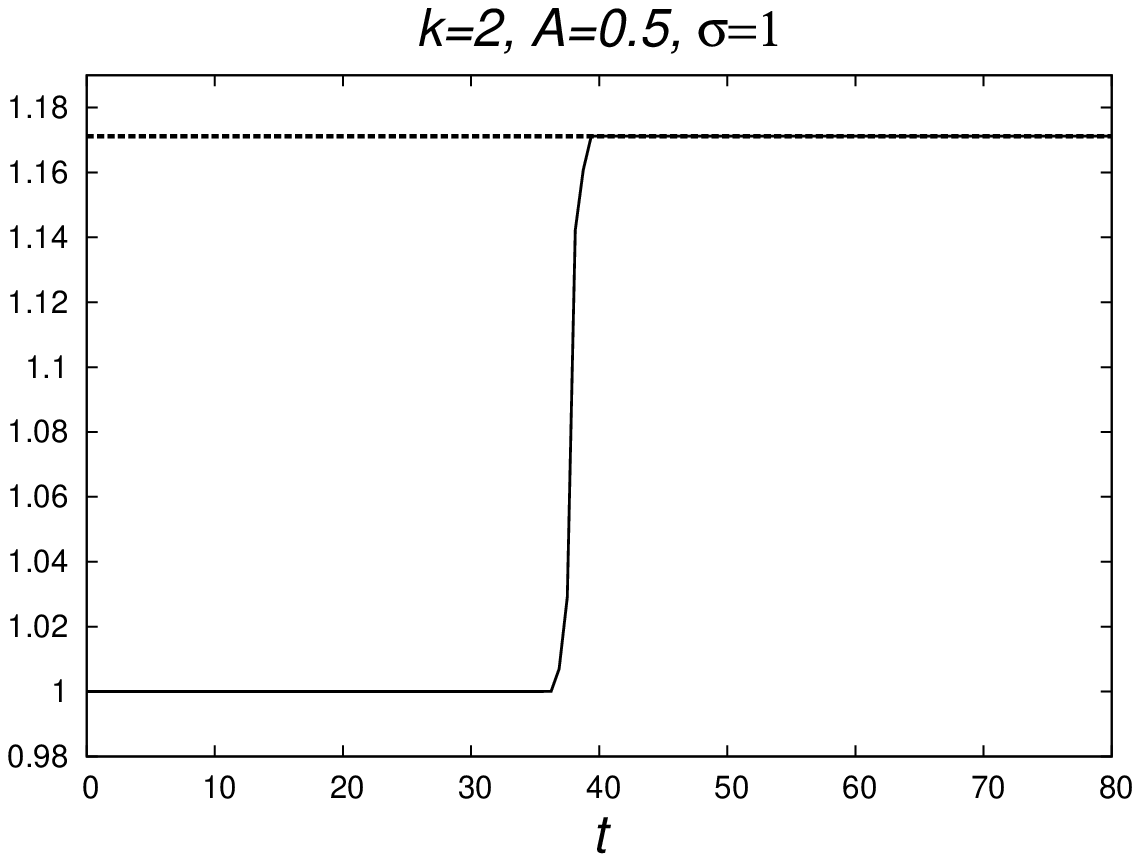}
\includegraphics[width=4.25cm]{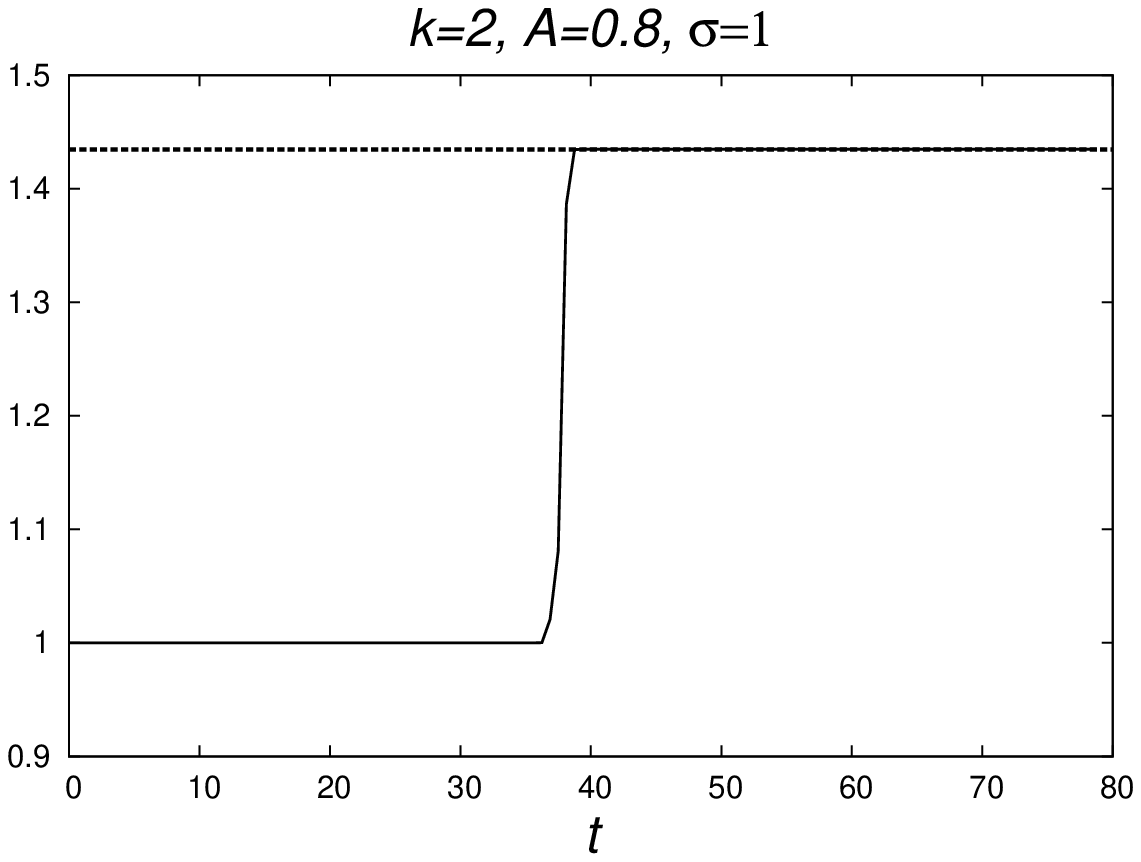}
\includegraphics[width=4.25cm]{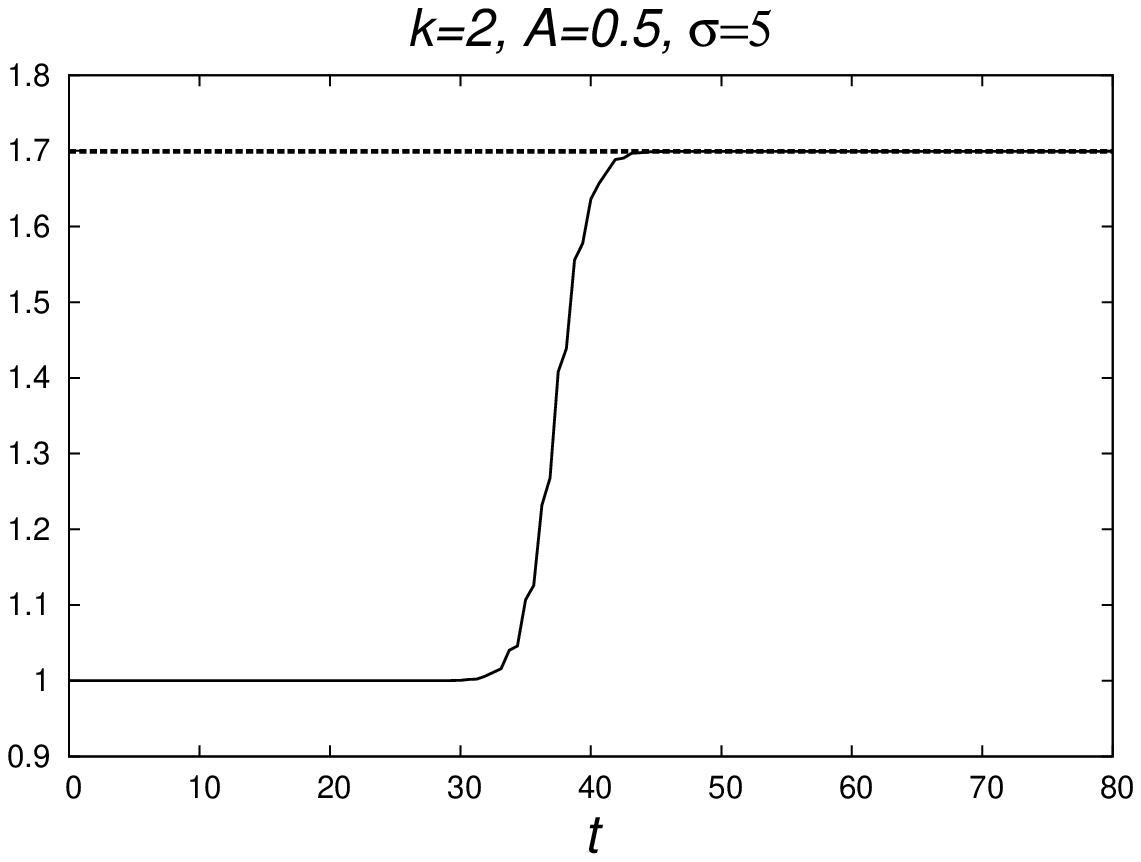}
\includegraphics[width=4.25cm]{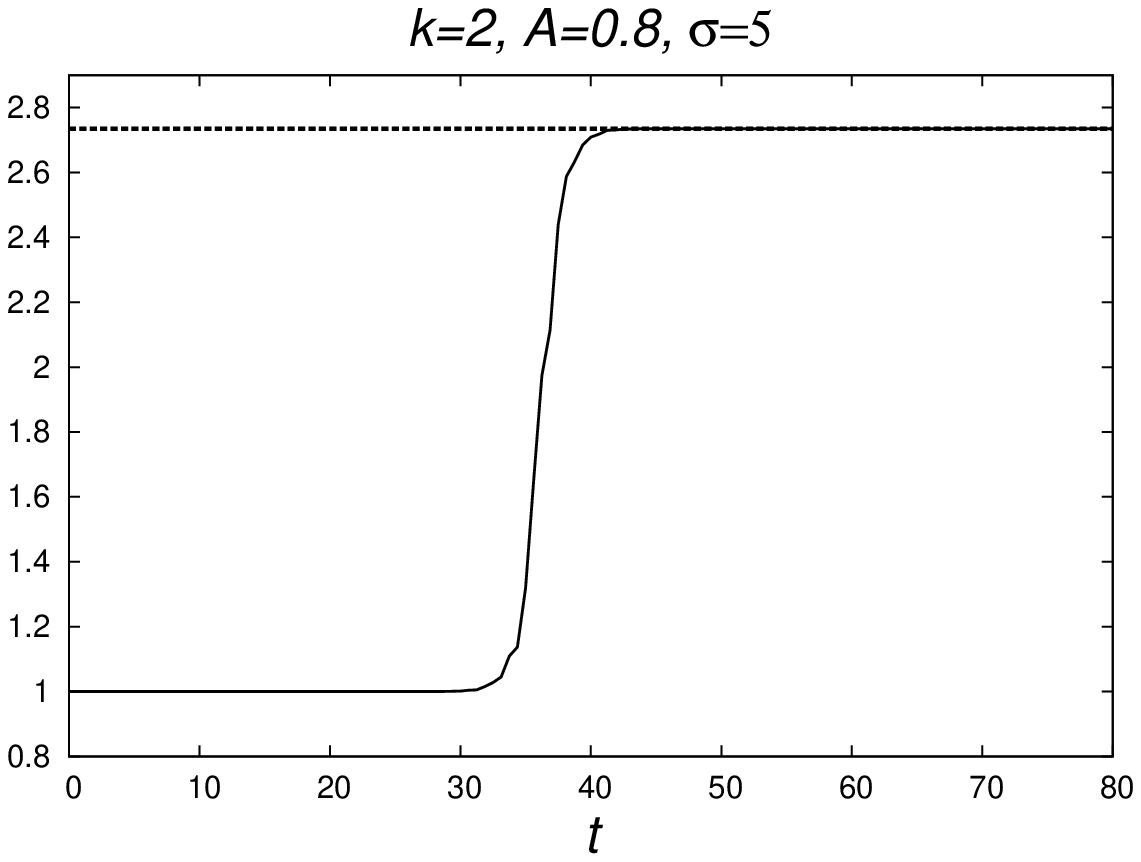}
\includegraphics[width=4.25cm]{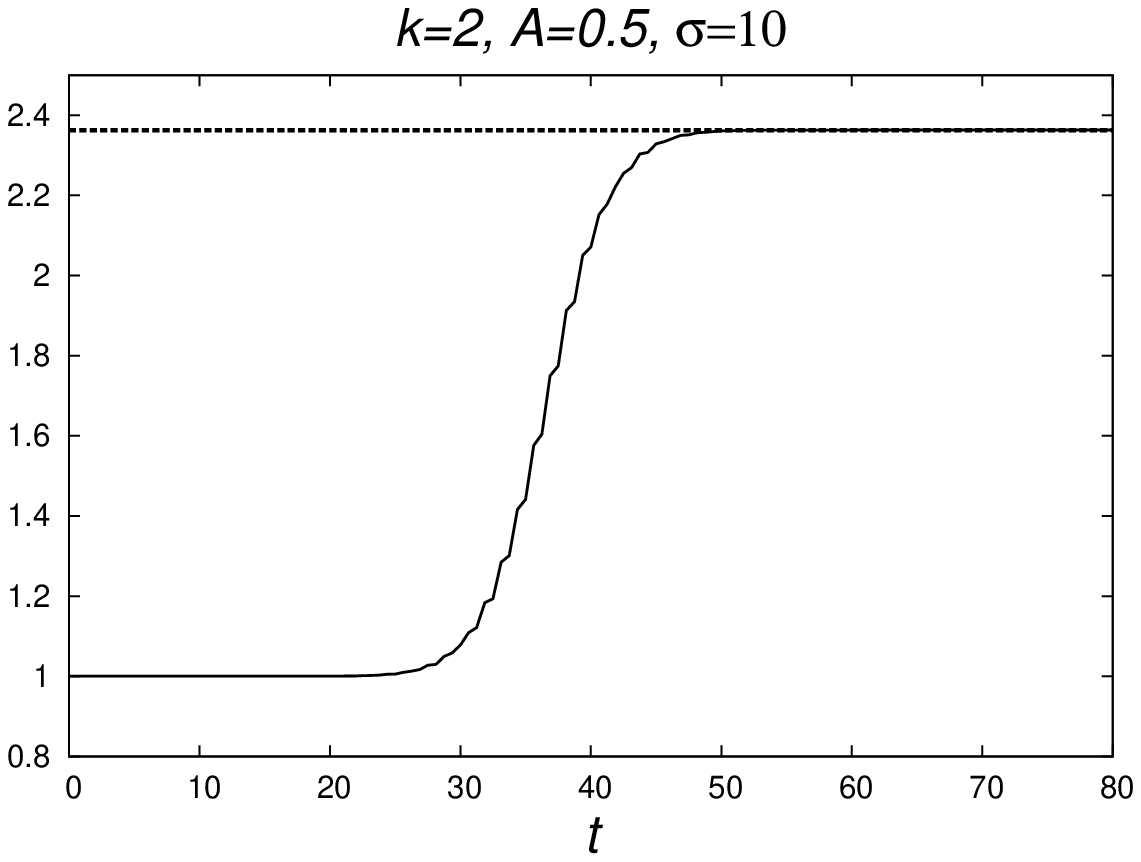}
\includegraphics[width=4.25cm]{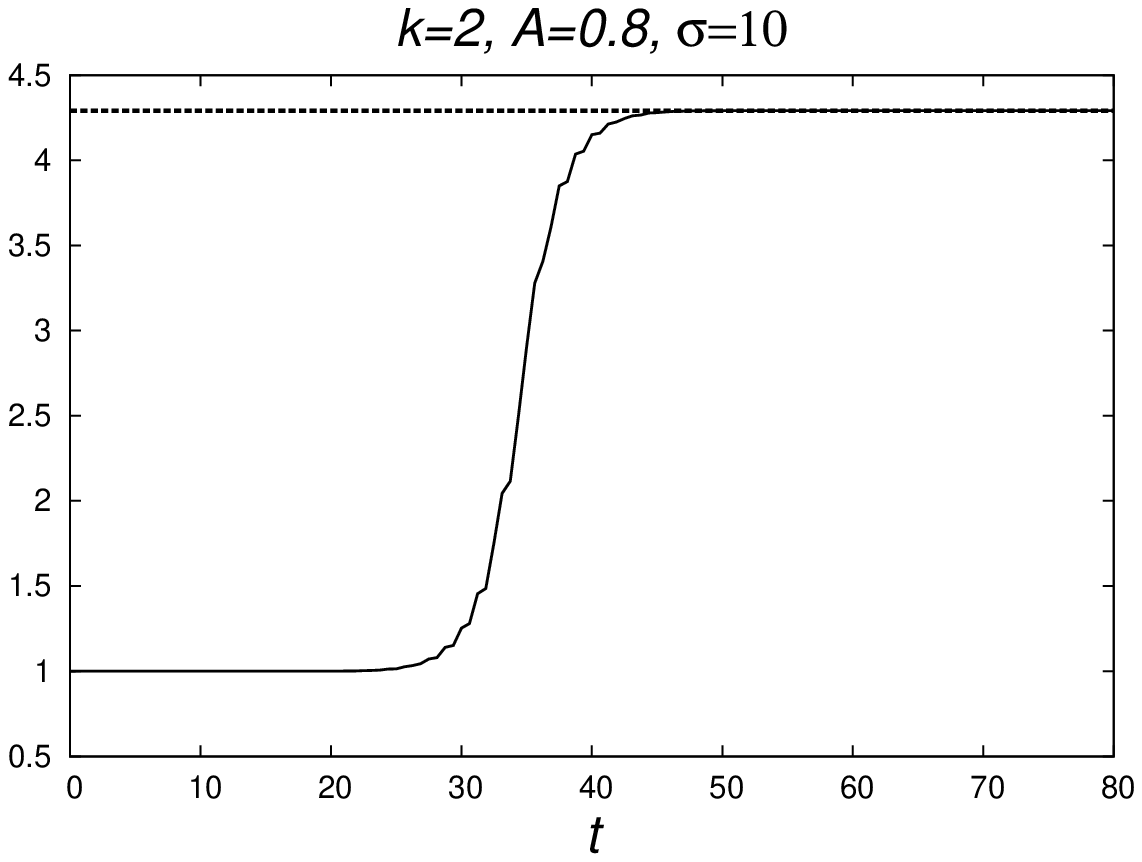}
\caption{\label{fig:k_2} We show the time evolution of the apparent horizon mass (continuous line) and the $M_{ADM}$ (dashed line) for the case $k=2$. Also full absorption was found in all cases.}
\end{figure}

Our results are shown in Figs. \ref{fig:k_0},  \ref{fig:k_0.5}, \ref{fig:k_1} and \ref{fig:k_2}, that correspond to the wave numbers $k=0,~k=0.5,~k=1$ and $k=2$ respectively, for the different values of $A$ and $\sigma$. In these figures we show the apparent horizon and $M_{ADM}$. These parameters are organized in such a way that we explore the suggestions in \cite{UrenaFernandez} related to the absorption of the scalar field in terms of the wave number, amplitude and width of the wave packet.
In all our runs we use $r_0=40$, which is far enough from the horizon.

Our first results confirm that for $k\gtrsim 1$ full absorption is observed independently of the value of $\sigma$ (Figs. \ref{fig:k_1}, \ref{fig:k_2}).

In the fixed space-time background case it was found that for values of $k$ smaller than 1 and big values of $\sigma$ not all the incoming scalar field was absorbed \cite{UrenaFernandez}. However as shown in Figs. \ref{fig:k_0}, \ref{fig:k_0.5} when $k=0.5$ full absorption was found in all cases and the only case where we find non-full (partial) absorption is that of $k=0$. Specifically for $\sigma < 2$, that is, smaller than the Schwarzschild radius, full absorption is found, whereas for $\sigma=5$ for the two values of $A$ we use, only $\sim 91\%$ of the ingoing scalar field is absorbed and for $\sigma=10$ we found that $\sim 65\%$ of the ingoing scalar field is absorbed. For comparison, in the fixed background case, it was found that the absorption could be as small as 50\% \cite{UrenaFernandez}.

\section{Discussion and conclusions}
\label{sec:conclusions}

We have tracked the spherically symmetric non-linear evolution of a massless scalar field being accreted into a Schwarzschild black hole. The aim of the work is to determine whether or not a scalar field is reflected or absorbed in terms of its wave number or wave packet width. As a first approximation we have considered the field to be massless.

With our results we generalize the analysis of the reflection and absorption of a massless scalar field in order to bound the possibility that black holes, like supermassive black holes could potentially absorb scalar fields that are being used as ingredients of nowadays cosmological models like scalar field dark matter and quintessence. Our generalization consists in adding the evolution of the geometry of space-time, unlike previous analyses where it remains fixed.

Our results confirm some of the previous predictions related to fixed background analyzes, specifically, we confirm that for spherical wave profiles of the scalar field with $k=0$ not all the incoming energy density is absorbed. Instead, we found the tendency to have less absorption when the initial wave packet width is bigger.

For all the other values of $k$ used, we found full absorption. Nevertheless, a detailed study would include other values of $k$ closer to 0.

In the context of the scalar field dark matter and dark energy models consisting of scalar fields, our results indicate that the evolution of the geometry allows the absorption of the total amount of the scalar field that is radially directed toward the black hole within a short time scale for large wave numbers, whereas, for $k=0$ a fraction of the scalar field remains outside the black hole.

At this point we have only performed our analyses using pretty general initial scalar field profiles. Nevertheless it would be interesting to study evolutions with quasi-stationary initial data (on a fixed background) like those proposed in \cite{unam} and study their evolution on an evolving space-time geometry.

Also, if specific models of dark matter or dark energy are to be studied, it is necessary to include the mass term of the scalar field and less restrictive symmetries, like rotation of the scalar field near the black hole and Kerr black holes.
 

\section*{Acknowledgments}

This research is partly supported by grants: 
CIC-UMSNH-4.9 and 
CONACyT 106466.
The runs were carried out in the IFM Cluster.



\begin{thebibliography}{10}

\bibitem{guth}
	A. H. Guth,
	Phys. Rev. {\bf D 23} (1981) 347.

\bibitem{quintessence}
	R.R. Caldwell, Rahul Dave, and P.J. Steinhardt, Phys. Rev. Lett. {\bf 80} (1988) 1582; Ivaylo Zlatev, Limin Wang and P.J. Steinhardt, Phys. Rev. Lett. {\bf 82} (1988) 896.

\bibitem{sfdm-gal} 
	F. S. Guzm\'an, T. Matos and H. Villegas-Brena, Astron. Nachr., {\bf 320} (1999) 97.
	F. S. Guzm\'an and T. Matos, Class. Quantum Grav., {\bf 17} (2000) L9-L16.
	T. Matos, F. S. Guzm\'an and D. N\'u\~nez. Phys. Rev. {\bf D 62} (2000) 061301(R).
	A. Arbey, J. Lesgourges and P. Salati, Phys. Rev. {\bf D 64} (2001) 123525.

\bibitem{sfdm-cosmo}
	V. Sahni, L. Wang, Phys. Rev. D 62, 103517 (2000).
	T. Matos and L. A. Ure\~na-L\'opez, 
	Class. quantum Grav. {\bf 17}, L75 (2000).

\bibitem{smbh}
	L. I. Caramete and P.L. Biermann, e-Print: arXiv: 1107.2244 [astro-ph.GA]. M. Volonteri and D. P. Stark, e-Print: arXiv:1107.1946 [astro-ph.CO]. L. Ferrarese, ApJ {578} (2002) 90.

\bibitem{seedBHs}
	D. J. Eisenstein and A. Loeb, ApJ {\bf 443} (1995) 11.
	S. M. Koushiappas, J. S. Bullock and A. Dekel, MNRAS {\bf 354} (2004) 292.

\bibitem{freitas} 
	S. Peirani and J. A. de Freitas-Pacheco, 
	Phys. Rev. {\bf D 77} (2008) 064023.

\bibitem{losecone} 
	A. P. Lightman and S. L. Shapiro,
         ApJ {\bf 211} (1977) 244-262.
         H. Zhao, M. G. Haehnelt, M. J. Rees,  
         New Astronomy, {\bf 7}, issue 7, (2002) 385-394.

\bibitem{Gilmore}
	J. I. Read and G. Gilmore, MNRAS {\bf 339} (2003) 949-956.

\bibitem{GuzmanLora2011}
	F. S. Guzm\'an, F. D. Lora-Clavijo, 
	Mon. Not. R. Astron. Soc. {\bf 415} (2011) 225-234.
	
\bibitem{GuzmanLora2011b}
	F. S. Guzm\'an, F. D. Lora-Clavijo, 
	Mon. Not. R. Astron. Soc. {\bf 416} (2011) 3083-3088.

\bibitem{LiddleUrena}
	L. Arturo Urena-Lopez, Andrew R. Liddle,
	Phys.Rev. D66 (2002) 083005.

\bibitem{CruzGuzmanLora}
	A. Cruz-Osorio, F. S. Guzman, F. D. Lora-Clavijo, JCAP 06 (2011) 029.

\bibitem{Anil}
	A. Zenginoglu,
	Class. Quantum Grav. 25:145002, 2008.
	A. Cruz-Osorio, A. Gonz\'alez-Ju\'arez, F. S. Guzm\'an, F. D. Lora-Clavijo,
	Rev. Mex. Fis. {\bf 56} (2010) 456-468.

\bibitem{UrenaFernandez}
	L. A. Ure\~na-L\'opez and L. M. Fern\'andez,
	Phys. Rev. D. {\bf 84} (2011) 044052.

\bibitem{mendozza}
	X. Hernandez, S. Mendoza, P. L. Rendon, C S. Lopez-Monsalvo, R. Velasco-Segura,
	Entropy {\bf 11} (2009) 17.

\bibitem{unam}
	J. Barranco et al.  
	Phys. Rev. D {\bf 84} (2011) 083008.

\bibitem{Harada}
	T. Harada, B. J. Carr
	Phys.Rev.D71:104010,2005.

\bibitem{oscillatons}
	M. Alcubierre, R. Becerril, F. S. Guzman, T. Matos, D. Nunez, L.A.Urena-Lopez,
	Class.Quant.Grav. 20 (2003) 2883-2904.

\bibitem{phantom}
	F. D. Lora-Clavijo, J. A. Gonzlaez, F S. Guzman,
	AIP Conf. Proc. {\bf 1256}, 339 (2010)

\bibitem{Jonathan}
	J. Thornburg, Phys.Rev.D59:104007,1999.
	J. Thornburg, e-Print: arXiv:gr-qc/9906022v2.
	R. L. Marsa, M. W. Choptuik,
	Phys.Rev.D54:4929-4943,1996.

\bibitem{excision}
	E. Seidel, W-M Suen, 
	Phys. Rev. Lett. {\bf 69} (1992) 1845.

\bibitem{mtw}
        Misner, Charles W., Thorne, Kip S., Wheeler, John A.
	Gravitation, W. H. Freeman and Company 1973.


\end{thebibliography}
\end{document}